\documentclass[12pt]{article}

\usepackage{times}
\usepackage{graphicx}
\usepackage{amsmath}

\usepackage[super]{natbib}
\title{Bending and Breaking of Stripes in a Charge-Ordered Manganite} 

\author
{Benjamin H. Savitzky,$^{1\dagger}$ Ismail El Baggari,$^{1\dagger}$ Alemayehu S. Admasu,$^{2}$\\ Jaewook Kim,$^{2}$ Sang-Wook Cheong,$^{2}$ Robert Hovden,$^{3\ddagger}$ \\and Lena F. Kourkoutis$^{3,4\ast}$\\
\\
\normalsize{$^{1}$Department of Physics, Cornell University, Ithaca, NY 14853, USA}\\
\normalsize{$^{2}$Rutgers Center for Emergent Materials and Department of Physics and Astronomy,}\\
\normalsize{Rutgers University, Piscataway, NJ 08854, USA}\\
\normalsize{$^{4}$School of Applied and Engineering Physics, Cornell University, Ithaca, NY 14853, USA}\\
\normalsize{$^{5}$Kavli Institute for Nanoscale Science, Cornell University, Ithaca, NY 14853, USA}\\
\\
\normalsize{$^\dagger$These authors contributed equally to this work.}\\
\normalsize{$^{\ddagger}$Current address: Dept. of Mat. Sci. \& Eng., University of Michigan, Ann Arbor, MI 48109, USA}\\
\normalsize{$^\ast$To whom correspondence should be addressed; E-mail:  lena.f.kourkoutis@cornell.edu.}
}

\date{}

\begin{document} 

\baselineskip24pt

\maketitle 

%%%%%%%%%%%%%%%%%%%%%%% Main Text %%%%%%%%%%%%%%%%%%%%%%%

\newpage

\section*{Abstract}

\textbf{
%Broad sentence
In complex electronic materials, coupling between electrons and the atomic lattice gives rise to remarkable phenomena, including colossal magnetoresistance and metal-insulator transitions.
%More specific background
Charge-ordered phases are a prototypical manifestation of charge-lattice coupling, in which the atomic lattice undergoes periodic lattice displacements (PLDs).
%Objective and Method
Here we directly map the picometer scale PLDs at individual atomic columns in the room temperature charge-ordered manganite Bi$_{0.35}$Sr$_{0.18}$Ca$_{0.47}$MnO$_{3}$ using aberration corrected scanning transmission electron microscopy (STEM).
%Result 
We measure transverse, displacive lattice modulations of the cations, distinct from existing manganite charge-order models.
%Result 
We reveal locally unidirectional striped PLD domains as small as $\sim$5 nm, despite apparent bidirectionality over larger length scales.
%Result
Further, we observe a direct link between disorder in one lattice modulation, in the form of dislocations and shear deformations, and nascent order in the perpendicular modulation.
%Broad conclusion
By examining the defects and symmetries of PLDs near the charge-ordering phase transition, we directly visualize the local competition underpinning spatial heterogeneity in a complex oxide.
}

\newpage

\section*{Introduction}

%%%%%%%%%%%%%%%%%%% Intro Paragraph %%%%%%%%%%%%%%%%%%%

Charge density wave (CDW) states are periodic modulations of both the electron density and atomic lattice positions. 
These states epitomize emergent order via electron-lattice interaction, and have taken a central role in understanding exotic phenomena in complex materials. 
CDWs mediate metal-insulator transitions, compete with high temperature superconductivity, and underlie the mechanism of colossal magnetoresistance in manganites \cite{Yoshida2014,Uehara1999,Tomioka1996,Wu2011,Chang2012,Sipos2008}.
Mounting evidence indicates that nanoscale spatial inhomogeneity between competing electronic phases plays a fundamental role in complex electronic systems quite broadly \cite{Dagotto2005,Milward2005,lang2002imaging}. 
For example, local competition and coexistence between charge-ordered and ferromagnetic regions is responsible for the colossal magnetoresistence effect in manganites, while
in cuprates, the suppression of superconducting order coincides with the emergence of charge-ordered patches \cite{Uehara1999,Hoffman2002,Wise2008}.
However, understanding of the microscopic mechanism driving such competition is lacking, requiring local interrogation of the atomic-scale behavior.

The manganese oxides provide a practical test bed for universal CDW phenomenology, as their strong electron-lattice coupling results in relatively robust charge and spin ordered phases \cite{Mourachkine2002}.
Striped states have been imaged in manganites with dark-field transmission electron microscopy (DF-TEM), however, resolution and signal-to-noise are limited in DF-TEM because electrons are collected from a small window of momentum space \cite{mori1998pairing,cox2008very}.
Moreover, the contrast mechanism of DF-TEM complicates interpretation, yielding inconsistent models of the modulation structure \cite{mori1998pairing,cox2008very,Louden2007}.
Therefore, atomically resolved measurements of PLDs have not been performed.

Here, we quantitatively map picometer-scale ($<$10 pm) periodic lattice displacements (PLDs) at individual atomic columns in the charge-ordered manganite Bi$_{0.35}$Sr$_{0.18}$Ca$_{0.47}$MnO$_{3}$ (BSCMO) near its transition temperature using scanning transmission electron microscopy (STEM).
In contrast to proposed manganite charge-order models \cite{mori1998pairing,goff2004,daoud2002,daoud2008,johnstone2012}, our data shows displacive, transverse, periodic modulations of the cation sites, with amplitudes of $\sim$6.2 pm/$\sim$8.2 pm on the A/B sites of the perovskite lattice.
We find two coexisting PLDs, forming locally unidirectional domains as small as $\sim$5 nm despite appearing bidirectional over larger length scales (see Fig.~\ref{F:FT}e), 
a distinction which is important but often challenging to establish \cite{cox2008very,Comin2015,Hoffman2002,Kajimoto2003}.
We unearth shear deformations and topological singularities in one PLD field, and establish that they coincide with nascent order in the perpendicular modulation.
Our results directly visualize the nanoscale complexity arising from competing phases and provide insight into the microscopic nature of charge-ordering \cite{Uehara1999,Dagotto2005,Milward2005,lang2002imaging,Hoffman2002,Comin2015}.

\section*{Results}

\begin{figure}[!htb]
  \includegraphics[width=0.5\linewidth]{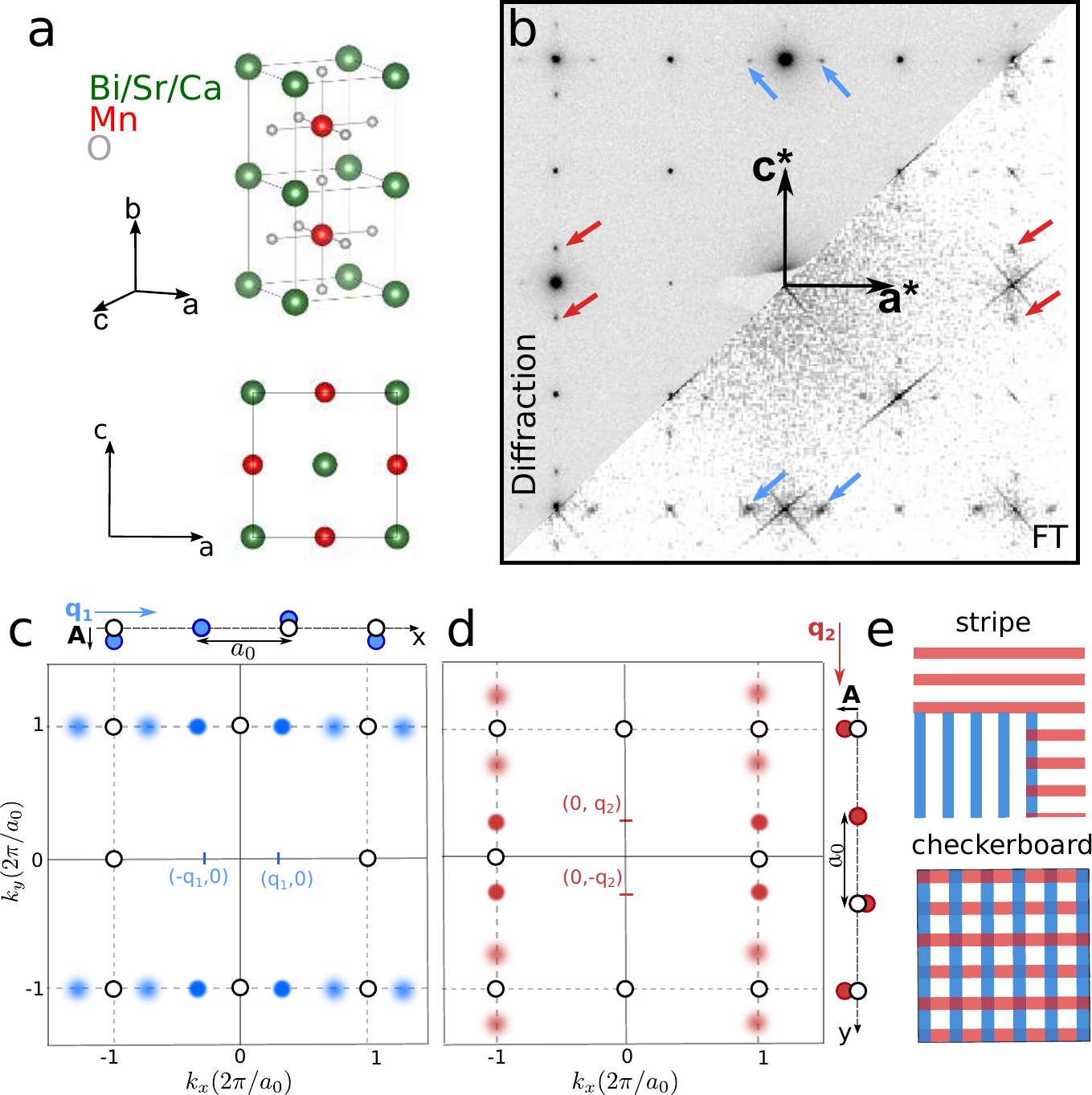}
  \caption{\textbf{Periodic lattice displacements in reciprocal space}
(\textbf{a}) The perovskite structure of Bi$_{0.35}$Sr$_{0.18}$Ca$_{0.47}$MnO$_{3}$ and the projection of the unit cell along the \textbf{b}-axis.
(\textbf{b}) Electron diffraction over a 1 $\mu$m selected area (\textit{upper left}) and the Fourier transform of a $\sim$30 nm field of view STEM image (\textit{lower right}) of BSCMO along the \textbf{b}-axis. Satellite peaks corresponding to two transverse and displacive modulations with perpendicular wavevectors $\mathbf{q_{1}} \approx 1/3 \ \mathbf{a^{*}}$  and $\mathbf{q_{2}} \approx 1/3 \ \mathbf{c^{*}}$ are indicated by blue and red arrows, respectively.
(\textbf{c},\textbf{d}) Schematic of the Fourier transform of a square lattice (for simplicity) displaced by transverse modulations along x and y, respectively. 
The intensity of a satellite peak is reduced when its reciprocal vector, $\mathbf{k}$ = ($k_{x},k_{y}$), is not parallel to the modulation polarization $\mathbf{A}_{i}$ and vanishes when $\mathbf{k}\cdot\mathbf{A}_i=0$.
(\textbf{e})  Stripe states contain locally unidirectional modulations, while checkerboard states contain overlapping bidirectional modulations.
Both stripe and checkerboard order are consistent with the reciprocal space data, which reflects the spatially averaged structure and cannot definitively determine the local symmetry.}
  \label{F:FT}
\end{figure}

\begin{figure}
  \includegraphics[width=\linewidth]{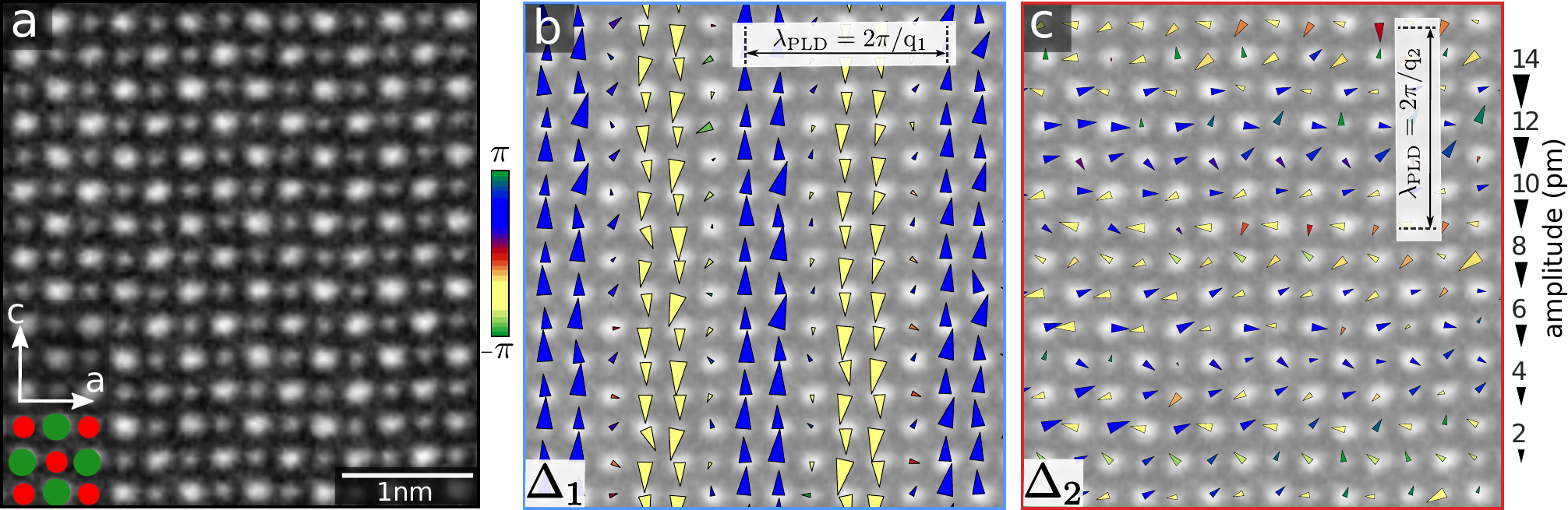}
  \caption{\textbf{Mapping picometer scale, periodic displacements of atomic lattice sites.} 
(\textbf{a}) HAADF-STEM projection image along the \textbf{b}-axis.  
The heavier (Bi, Sr, Ca)-sites (green) appear brighter than the lighter Mn-sites (red).
(\textbf{b}) Mapping picometer scale displacements $\mathbf{\Delta}_1(\mathbf{r})$ at each atomic lattice site in response to a single modulation wavevector $\mathbf{q_{1}}$. 
PLD maps indicate a displacive modulation rather than an intensity modulation (cation order, charge disproportionation) with transverse polarization and 3$a$ periodicity. 
Triangles represent displacements, with the area scaling linearly with displacement amplitude.
The color represents the angle of the polarization vector, $\mathbf{A}_1$, relative to $\mathbf{q}_1$ where blue (yellow) correspond to 90$^\circ$(-90$^\circ$). 
(\textbf{c}) Map of $\mathbf{\Delta}_2(\mathbf{r})$ displacements at each atomic lattice site in response to $\mathbf{q}_2$ in the same region as (a,b).  The significantly weaker $\mathbf{\Delta}_2(\mathbf{r})$ response is characteristic of locally striped, rather than checkerboard, ordering.}
  \label{F:PLD}
\end{figure}

The BSCMO orthorhombic perovskite lattice (space group $Pnma$, Fig.~\ref{F:FT}a) is imaged in projection along the b-axis with aberration-corrected high-angle annular dark-field (HAADF)-STEM (Fig.~\ref{F:PLD}a), which is sensitive to the Coulomb potential of the atomic nuclei;
heavier Bi/Sr/Ca atomic columns (A-sites) appear brighter than lighter Mn columns (B-sites) in the $Z$-contrast image.
Electron diffraction (Fig.~\ref{F:FT}b, \textit{upper left}) shows a constellation of satellite peaks indicating two transverse, displacive PLDs (Figs.~\ref{F:FT}c,d) offsetting the atomic lattice with displacements
\begin{equation}
\label{E:modulation_field}
\mathbf{\Delta}_{i}(\mathbf{r}) = \mathbf{A}_{i}\sin\left( \mathbf{q}_i\cdot\mathbf{r} + \phi_{i} \right), \qquad i\in\left\{1,2\right\}
\end{equation}
where $\mathbf{A}_i$, $\mathbf{q}_i$, and $\phi_i$ are the PLD amplitude vector, wavevector, and phase, respectively, and $\lvert\mathbf{q}_i\rvert\approx\frac{1}{3}$ reciprocal lattice units.
Diffraction shows coexistence of the two orthogonal PLDs within a 1 $\mu$m selected area.
A STEM Fourier transform (Fig.~\ref{F:FT}b, \textit{lower right}) shows coexistence within a $\sim$30 nm field of view.
In order to further investigate the local PLD structure, we extract the displacement vectors associated with each of the two modulations at every atomic site to generate the PLD maps shown in Figs.~\ref{F:PLD}b,c.

To calculate the PLD fields $\mathbf{\Delta}_i(\mathbf{r})$ shown in Fig.~\ref{F:PLD}, we first fit all atomic positions in our STEM data with $\sim$2 picometer precision, an approach which has recently emerged as a powerful, quantitative characterization tool \cite{Nelson2011,Yankovich2014,Yadav2016}.
However, in contrast to prior STEM atom tracking work, the key challenge in mapping PLDs is defining an appropriate reference lattice, which is complicated by the presence of local PLD phase variations and multiple interpenetrating modulations.
Our approach generates a reference image in which the contribution of a single modulation has been selectively removed, by damping all of the relevant satellite peaks from the Fourier transform of the original image.
Fitting and subtracting corresponding lattice positions from the image pair yields $\mathbf{\Delta}_{i}(\mathbf{r})$ quantitatively.
Damping the $\mathbf{q}_1$ satellite peaks (Figs.~\ref{F:FT}b,c, \textit{blue arrows}) generates a map of $\mathbf{\Delta}_1(\mathbf{r})$ (Fig.~\ref{F:PLD}b), while damping the $\mathbf{q}_2$ satellite peaks (Figs.~\ref{F:FT}b,d, \textit{red arrows}) maps $\mathbf{\Delta}_2(\mathbf{r})$ (Fig.~\ref{F:PLD}c).
Simulations indicate that our method accurately reconstructs the PLD structure everywhere except at lattice sites directly adjacent to atomically sharp discontinuities in the PLD field.
Analytical and algorithmic details, simulations, and error analysis are found in Supplemental Text and Supplementary Figs.~S4-11.

The microscopic structure of charge-ordered phases in manganites remains contested\cite{goff2004,daoud2002,daoud2008,johnstone2012};
here, the $\mathbf{\Delta}_1(\mathbf{r})$ map in Fig.~\ref{F:PLD}b furnishes real-space evidence for displacive lattice modulations of both the Bi/Sr/Ca sites and the Mn sites, with respective amplitudes of $\sim$6.2 pm and $\sim$8.2 pm on the maximal sites (see Supplementary Fig.~S12).
The displacements are transverse to the modulation wavevector and generate a tripled unit cell.
The historically prevailing model conjectures the localization and ordering of Mn$^{3+}$-Mn$^{4+}$ ions, which in turn activates an alternating compression and expansion of oxygen octehedra (Jahn-Teller effect)\cite{mori1998pairing}.
Other works propose the formation of Mn pairs (Zener polarons) with minimal valence modulations\cite{daoud2002,Louden2007}.
Our data suggests a different model.
The strong structural modulation shown in Fig.~\ref{F:PLD}b is consistent with the softening of a phonon mode, and the pattern of displacements provides a structural model to further investigate the microscopic origin of the modulated state.

The superposition of multiple modulations can further mask the underlying microscopic mechanism behind PLD formation. 
For instance, distinguishing overlapping modulations \linebreak (checkerboards) from spatially anti-correlated unidirectional domains (stripes) is essential but challenging, as both have the same spatially averaged symmetry (Fig.~\ref{F:FT}b--d) \cite{cox2008very,Comin2015,Kajimoto2003,Robertson2006,DelMaestro2006}.
Our data clearly indicates that locally, BSCMO forms striped states: where one PLD is suppressed, the other is strong, starkly illustrated in the $\mathbf{\Delta}_1(\mathbf{r})$ and $\mathbf{\Delta}_2(\mathbf{r})$ maps of identical regions in Figs.~\ref{F:PLD}b,c.

\begin{figure}
  \begin{minipage}[b]{0.65\textwidth}
    \includegraphics[width=\textwidth]{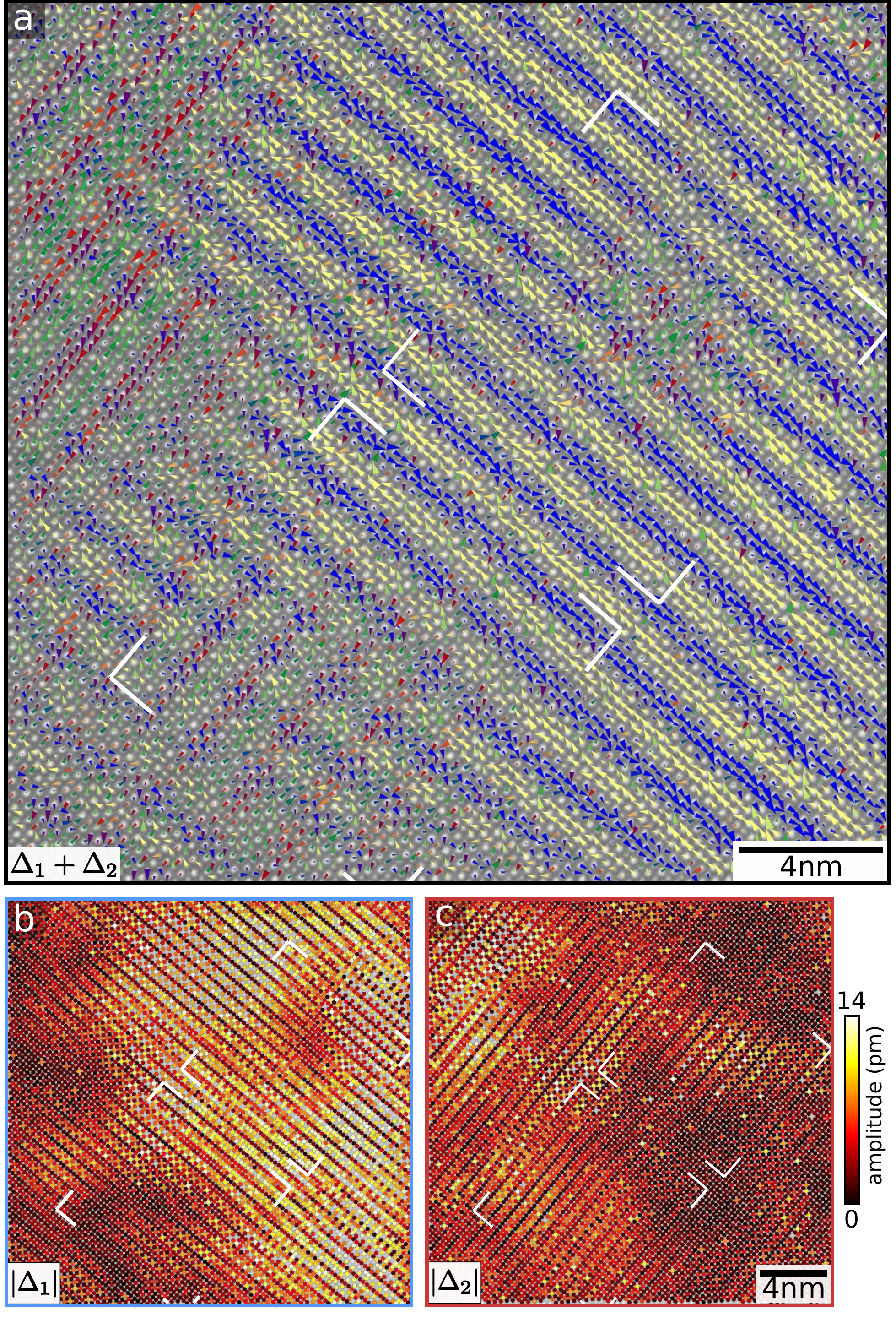}
  \end{minipage}\hfill
  \begin{minipage}[b]{0.32\textwidth}
    \caption{\textbf{Nanoscale domain structure and local symmetry of PLD stripes.}  
(\textbf{a}) Combined PLD map showing the displacements $\mathbf{\Delta}(\mathbf{r}) = \mathbf{\Delta}_1(\mathbf{r})+\mathbf{\Delta}_2(\mathbf{r})$ at all $\sim$9,000 atomic sites in the $\sim$30 nm field of view.  
Colors indicates the polarizations relative to $\mathbf{q}_1$ as in Fig.~\ref{F:PLD}, and triangle areas scale linearly with the displacement magnitude.
(\textbf{b},\textbf{c}) Maps of the magnitudes $\lvert\mathbf{\Delta}_1(\mathbf{r})\rvert$ and $\lvert\mathbf{\Delta}_2(\mathbf{r})\rvert$ of the displacements due to each PLD individually reveals that the two PLD strengths are anticorrelated: when one is strong, the other is weak. 
The PLDs are stripe ordered, segregated into nanoscopic domains.
The two indicated regions (\textit{white corners}) are further analyzed in Fig.~\ref{F:Defects}.}
    \label{F:Domains}
  \end{minipage}
\end{figure}

Zooming out, Fig.~\ref{F:Domains}a maps the combined displacement field $\mathbf{\Delta}(\mathbf{r})=\mathbf{\Delta}_1(\mathbf{r})+\mathbf{\Delta}_2(\mathbf{r})$ over a $\sim$30 nm field of view, in which
a $\mathbf{\Delta}_1$--dominant region, readily identified by its transverse polarization relative to $\mathbf{q}_1$ (\textit{blue/yellow triangles}), occupies the right side of the frame, while a $\mathbf{\Delta}_2(\mathbf{r})$--dominant region occupies the upper left corner (\textit{red/green triangles}).
Mapping the displacement magnitudes $\lvert\mathbf{\Delta}_1(\mathbf{r})\rvert$ and $\lvert\mathbf{\Delta}_2(\mathbf{r})\rvert$ visualizes the striped domain structure, revealing complex domain morphology with islands of strong modulations (6-11pm) and basins of PLD suppression (0-3pm) (Fig.~\ref{F:Domains}b,c).
Notably, regions in which both $\mathbf{\Delta}_1(\mathbf{r})$ and $\mathbf{\Delta}_2(\mathbf{r})$ are present are also observed, such as the bottom left corner of Figs.~\ref{F:Domains}a--c.
Quenched disorder tends to broaden phase transitions and favors enhanced isotropy in the nascent ordered state, and theoretically has been shown to induce apparent fourfold symmetry in 2D striped phases \cite{Robertson2006,DelMaestro2006,LeTacon2013}.
We believe the checkerboard-like regions we observe may result from quenched disorder; 
varying intensity of atomic columns clearly indicates frozen cation disorder in our data (see Supplementary Fig.~S14).
Alternatively, checkerboard-like ordering could result from projection through stacked $\mathbf{\Delta}_1(\mathbf{r})$ and $\mathbf{\Delta}_2(\mathbf{r})$ domains in the out-of-plane (b-axis) direction.
In either case, the two modulations are predominantly anti-correlated in our data, and we conclude that the symmetry breaking in the disorder-free ``clean'' limit in this system is very likely striped.

\begin{figure}
  \includegraphics[width=0.9\linewidth]{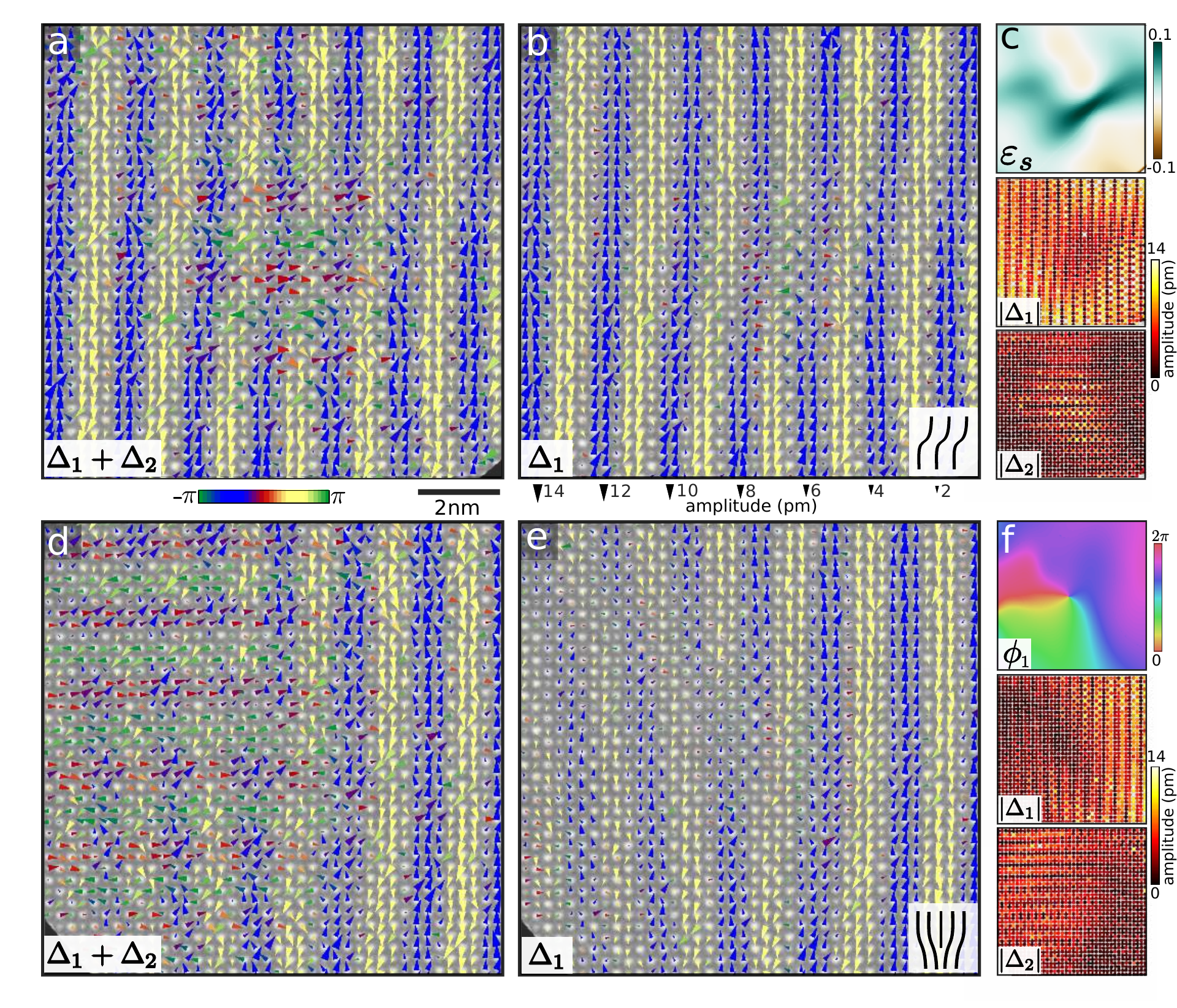}
  \caption{\textbf{Nascent order coincident with PLD defects} (\textbf{a}) A complete $\mathbf{\Delta}=\mathbf{\Delta}_1 + \mathbf{\Delta}_2$ map of a $\sim$5 nm region of incipient $\mathbf{\Delta}_2$ order, and a coinciding shearing of the $\mathbf{\Delta}_1$ modulation. (\textbf{b}) A $\mathbf{\Delta}_1$ map of the same region highlights the bending wavefronts, and reveals attenuation of the PLD amplitude and some rotation of the displacement vectors in the defective region.  (\textbf{c})  The shear strain $\varepsilon_s$, $\lvert\mathbf{\Delta}_1\rvert$, and $\lvert\mathbf{\Delta}_2\rvert$ (\textit{top}, \textit{middle}, \textit{bottom}, respectively) in the same region.  The maximal shearing aligns with attenuation of $\mathbf{\Delta}_1$ and emergence of $\mathbf{\Delta}_2$.
(\textbf{d})  A complete $\mathbf{\Delta}=\mathbf{\Delta}_1 + \mathbf{\Delta}_2$ map of the interface between a $\mathbf{\Delta}_1$-dominant region and coexisting $\mathbf{\Delta}_1$ and $\mathbf{\Delta}_2$ order.  (\textbf{e}) A $\mathbf{\Delta}_1$ map of the same region reveals a dislocation in $\mathbf{\Delta}_1$, with a burgers vector of $\lambda_{PLD}\widehat{\mathbf{q}_1}$.  Analogous to the elastic deformation of an atomic lattice about crystal dislocation, the elastic response of the PLD includes bending and compression of wavefronts and local displacement rotations.  Some attenuation of $\mathbf{\Delta}_1$ is apparent in the mixed region.  (\textbf{f})  The phase $\phi_1$, $\lvert\mathbf{\Delta}_1\rvert$, and $\lvert\mathbf{\Delta}_2\rvert$ (\textit{top}, \textit{middle}, \textit{bottom}, respectively) in the same region.  $\mathbf{\Delta}_1$ weakens and $\mathbf{\Delta}_2$ grows within $\sim\lambda_{PLD}$ of the defect core, where $\phi_1$ exhibits an expected $2\pi$ winding.  A narrow inlet of $\lvert\mathbf{\Delta}_1\rvert$ amplitude collapse extends from the upper left to the singularity.
  }
	\label{F:Defects}
\end{figure}

CDW domain nucleation near T$_c$ remains a poorly understood process, particularly in the presence of disorder \cite{LeTacon2013,Liu1998}.
We observe PLD defects coincident with both domain boundaries and nascent domain structures, suggesting their involvement in mediating domain growth and termination.
Figures~\ref{F:Defects}a--c magnify the region containing a $\sim$5 nm island of $\mathbf{\Delta}_2$ order embedded in a $\mathbf{\Delta}_1$ domain (Figs.~\ref{F:Domains}c--e, \textit{upper white delimiters}).
Inspection of the $\mathbf{\Delta}_1+\mathbf{\Delta}_2$ map  (Fig.~\ref{F:Defects}a) reveals shearing in $\mathbf{\Delta}_1$ as it passes through the $\mathbf{\Delta}_2$ island, evident in the offset of the wavefronts by $\sim$2 atomic rows. 
Mapping $\mathbf{\Delta}_1$ only (Fig.~\ref{F:Defects}b) accentuates the shear deformation, and exposes $\mathbf{\Delta}_1$ attenuation in the strained region, along with rotation of the displacement vectors to roughly align with the local wavefront orientation.
To quantify these observations we map the elastic shear strain field, $\varepsilon_s(\mathbf{r})$, reflecting local bending in the $\mathbf{\Delta}_1$ PLD, along with the magnitudes of the two modulations $\lvert\mathbf{\Delta}_1\rvert$ and $\lvert\mathbf{\Delta}_2\rvert$ (Fig.~\ref{F:Defects}c, \textit{top}, \textit{middle}, \textit{bottom}, respectively).
$\varepsilon_s(\mathbf{r})$ is calculated by extracting the local PLD phase ($\phi\to\phi(\mathbf{r})$ in Eq.~\ref{E:modulation_field}) \cite{Lawler2010} then computing $\varepsilon_s(\mathbf{r}) = \frac{1}{2}\frac{\widehat{\mathbf{q}_\perp}}{\lvert\mathbf{q}\rvert} \cdot\mathbf{\nabla}\phi(\mathbf{r})$ (see Supplementary Text) \cite{Feinberg1988,Brazovskii2004}.
The shear defect plainly coincides with abatement of $\mathbf{\Delta}_1$, and strengthening of $\mathbf{\Delta}_2$.

Figures~\ref{F:Defects}d--f magnify a domain boundary (Fig.~\ref{F:Domains}c--e, \textit{lower white delimiters}).
Exclusive $\mathbf{\Delta}_1$ order occupies the right side of the frame in Fig.~\ref{F:Defects}d, while the displacements to the left suggest an intricate interweaving of two the modulations.
Mapping $\mathbf{\Delta}_1$ only (Fig.~\ref{F:Defects}e) reveals a prominent dislocation in the PLD, in which a single wavefront abruptly terminates.
Analogous to edge dislocations in crystalline solids, where the abrupt termination of a row of atoms is accompanied by elastic deformation in the surrounding lattice, we observe elastic deformation of the PLD about the singularity, evident in the warped wavefronts flanking the dislocation core.
No defects in the underlying lattice are observed (see Supplementary Fig.~S14), and the PLD phase $\phi(\mathbf{r})$ exhibits an expected $2\pi$ winding about the discontinuity (Fig.~\ref{F:Defects}f, \textit{top}).
The interface between the $\mathbf{\Delta}_1$ dominant domain and the mixed region occurs within a single PLD wavelength of the defect core, as once again disorder in one modulation accompanies commencement of order in the other.
Maps of the PLD magnitudes $\lvert\mathbf{\Delta}_1\rvert$ and $\lvert\mathbf{\Delta}_2\rvert$ (Fig.~\ref{F:Defects}f, \textit{middle}, \textit{bottom}, respectively) reinforce these observations.
Moreover, theory predicts modulation amplitude collapse at singularities to prevent divergence of the energy density, and the $\lvert\mathbf{\Delta}_1\rvert$-map exhibits a narrow inlet of collapsed amplitude extending from the upper left to the defect core, suggesting complex domain restructuring to accommodate the high energy feature \cite{Feinberg1988,Brazovskii2004,Lee1979,coppersmith1991diverging}.
While displacements at atomic sites directly adjacent to a true singularity will not be accurately reconstructed, we believe the displacements extracted by our method are valid everywhere, because damping and distortion in the defect's central region yields reasonably smooth variations of the displacements (see Supplementary Text and Supplementary Figs. S6, S7, and S9).

In general, many factors appear to govern macroscopic behavior in complex electronic systems.
The nanometer-scale interplay between new order and defects in an extant order parameter may be one ubiquitous element, as in emergent charge-ordered states at the core of superconducting vortices, emergent ferromagnetic or superconducting order at CDW discommensuration domain boundaries, or competing PLD domains \cite{Sipos2008,Milward2005,Hoffman2002}.
The picture is further complicated by the presence of quenched impurities which can pin defects, stabilize ordered phases above T$_c$, or lead to complex mixed phases, and may play a role in the phenomena we observe \cite{Dagotto2005,LeTacon2013,Lee1979}.
Even more fundamental, and still elusive, is a microscopic understanding of which couplings give rise to which competing states, and how.
In addition to providing a new structural model of charge-ordered manganites, our data renders the interacting order and disorder in competing PLDs immediately visually apparent: where one modulation bends or `breaks', the other manifests.
These first observations of the atomically resolved structure of a PLD suggest new lines of inquiry into the nature of modulated phases.

%%%%%%%%%%%%%%%%%%%%   Figures   %%%%%%%%%%%%%%%%%%%%%%%

% Figure 1: PLDs in fourier space
\clearpage
\section*{Figures}

% Figure 2: PLDs in real space

% Figure 3: PLD domain structure

% Figure 4: PLD defects and nascent order

%%%%%%%%%%%%%%%%%%%%% Methods %%%%%%%%%%%%%%%%%%%%%%%%%

\clearpage

\section*{Methods}

Bi$_{1-x}$Sr$_{x-y}$Ca$_{y}$MnO$_{3}$ (BSCMO) single crystals were grown using the flux method, using Bi$_2$O$_3$, CaCO$_3$, SrCO$_3$, and Mn$_2$O$_3$.
Temperature-dependent electrical resistivity measurements (Supplementary Fig.~S1) show a transition at $\sim$300 K, which is associated with the onset of charge order. 
Sample preparation for electron microscopy and energy dispersive X-ray spectroscopy (EDX) were performed on a FEI Strata 400 Focused Ion Beam (FIB). 
From EDX, the composition was determined to be approximately $x=0.65$ and $y=0.47$ (Supplementary Fig.~S2) with negligible variations over the whole sample (size 0.34 $\times$ 0.28 mm). 

A thin, electron transparent cross section of BSCMO was extracted using FIB lift out, with estimated thickness in the imaging region ranging from 10 to 30 nm.
Based on electron diffraction, the orientation of the sample was along the \textbf{b} direction (orthorhombic axis) in the $Pnma$ space group (Supplementary Fig.~S3).
At room temperature (293K), BSCMO exhibits satellite peaks (\textit{arrows}), indicating the presence of charge ordering.

We performed atomic-resolution imaging in an aberration corrected scanning transmission electron microscope (FEI Titan Themis) operating at 300 kV. 
The beam convergence angle was 30 mrad. 
For Z-contrast imaging, we collected high-angle annular dark field images where the inner and outer collection angles were 68 and 340 mrad, respectively.
During STEM imaging the sample experienced a $\sim$2 Tesla magnetic field due to its position inside the objective lens, as determined from a Hall bar measurement. 
In order to minimize the effect of scan noise and stage drift, we acquired 20 to 30 images in rapid succession with a 2$\mu$s/pixel dwell time.
We registered and averaged stacks of images using both rigid registration and non-rigid registration methods and found similar results.
Data was acquired at 27.4 pm/pixel, and acquisition was optimized for pixel density, field of view and Fourier space sampling.
We performed atom-tracking with $\sim$2 pm precision (Supplementary Fig.~S11 and Supplementary Text) by fitting two-dimensional Gaussians to atomic columns using various optimization packages (scipy, photutils, MATLAB) and found consistent results.

%%%%%%%%%%%%%%%%%%%%%   Bibliography   %%%%%%%%%%%%%%%%%%%%
\clearpage

%%%% Bibliography with all author listed %%%%

%%%%%%%%%%%%% Acknowledgements %%%%%%%%%%%%%%%%
%%%%%%%%%%%% Author Contributions %%%%%%%%%%%%%
%%%%%%%%%%%% Competing Interests %%%%%%%%%%%%%%
%%%%%%%% Correspondence and Materials %%%%%%%%%

\clearpage

\section*{Acknowledgements}
We thank Michael J. Zachman and David J. Baek for experimental support.
We acknowledge support by the Department of Defense Air Force Office of Scientific Research (FA 9550-16-1-0305) and the Packard Foundation. 
The FEI Titan Themis 300 was acquired through NSF-MRI-1429155, with additional support from Cornell University, the Weill Institute and the Kavli Institute at Cornell. 
This work made use of the Cornell Center for Materials Research facilities supported through the NSF MRSEC program (DMR-1120296). 
B.H.S. was supported by NSF GRFP grant DGE-1144153.  
The work at Rutgers was supported by the Gordon and Betty Moore Foundation’s EPiQS Initiative through Grant GBMF4413 to the Rutgers Center for Emergent Materials.

\section*{Author Contributions}
A.S.A., J.K., and S-W.C. synthesized the crystals and performed electrical transport characterization.
B.H.S., I.E., R.H, and L.F.K acquired and analyzed STEM and electron diffraction data.
B.H.S., I.E., R.H., and L.F.K. wrote the manuscript.
The work was conceived and guided by L.F.K.
All authors discussed results and commented on the manuscript.

\section*{Competing Interests Statement}
The authors declare that they have no competing financial interests.

\section*{Materials \& Correspondence} 
Corresponsence and requests for materials should be addressed to L.F.K. \newline (lena.f.kourkoutis@cornell.edu).

\end{document}

\maketitle
\clearpage

% The \section*{} command should be used for "Major Heads" - i.e. 
% Materials and Methods, Supplementary Text, and each supplementary figure.
% The \subsection*{} command should be used for "Minor heads"

\section*{Supplementary Text}

\subsection*{Periodic Lattice Displacement Mapping Method}

\subsubsection*{Overview}

Extracting local atomic positions with picometer precision is a well established and powerful tool for analysis of high resolution STEM data, and has been used to great effect in describing, for example, local polarization in ferroelectrics, interfacial coupling in oxide heterostructures, and tunable octahedral rotations\cite{Jia2008,Nelson2011,Catalan2011,Tang2014,Yadav2016,Moon2014,Kan2016,Liao2016}.
However, the ability to measure the positions of atomic nuclei is a necessary but not sufficient condition to locally map periodic lattice displacements.
The key challenge is defining a reference lattice: in order to calculate the atomic displacements, each atomic position measured in the raw data must be compared to some suitable reference position.  
In ferroelectrics, defining a reference is comparatively straightforward, for example by measuring the displacements of a central B-site atom with respect to a surrounding A-site cage or an oxygen cage in a unit cell of ABO$_3$\cite{Jia2008,Nelson2011,Catalan2011,Tang2014,Yadav2016}.
Here, in contrast, there exists no simple reference against which to measure the displacement of a given atomic site, further exacerbated by the possibility of disorder, distinct sublattice behavior, and multiple modulation wavevectors.

The method used in this report defines a reference lattice against which the displacements of interest can be measured by leveraging the convenient decoupling of the PLD from its underlying unmodulated lattice in Fourier space: the unmodulated lattice appears in Fourier space as the usual Bragg peaks, and the PLD as satellite peaks decorating each Bragg peak.
By carefully damping the satellite peaks in Fourier space, the contribution of the PLD only is removed, and a reference lattice can be extracted.
In addition to making local PLD mapping tenable in the first place, this approach has two notable advantages:
first, multiple independent modulations can be individually extracted and mapped because they are decoupled in Fourier space; and second, local distortions from the imaging process are naturally accounted for because they are present in both the original and reference images.

Below, the details of the method are described.
First, the data processing procedure itself is presented.  
Next, the validity of the approach is confirmed by comparison to lattice displacements calculated using a coarser approach which involves only real space measurements from the raw data.  
The limits and regime of validity of the approach are then discussed and illustrated using a variety of simulated datasets, and rigorous interpretation of the extracted displacements is addressed.  
The importance of judiciously choosing an appropriate Fourier space mask is then discussed.  
Finally, we discuss important analytical details of our approach relating PLDs in real space and Fourier space.

\subsubsection*{Data Processing}

\begin{figure}
  \includegraphics[width=6in]{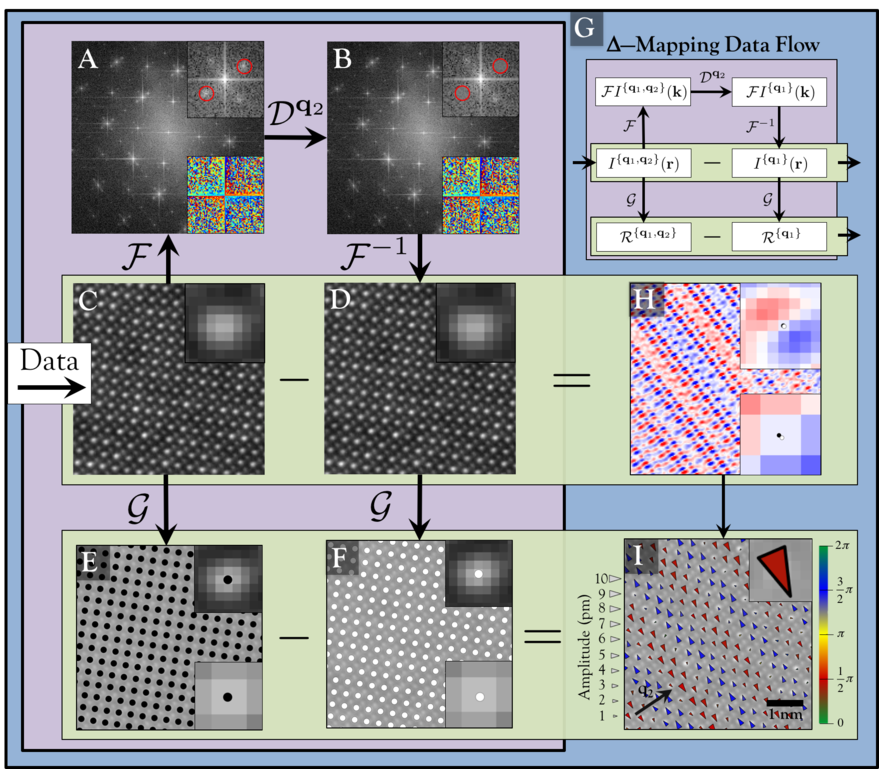}
  \caption{\textbf{PLD map data processing} 
PLD maps are generated beginning with a single high resolution STEM image, obtained by cross correlating and averaging $\sim$20-30 fast scanned images (C).  
A Fourier transform is calculated (A), and all satellite peaks corresponding to a modulation wavevector of interest are extracted (\textit{upper inset}).  
These satellite peaks are damped to the background level (B, \textit{upper inset}) while the phase is left unchanged (A,B, \textit{lower insets}).
Taking an inverse Fourier transform yields a reference lattice (D).
The difference between the original and reference image qualitatively depict the PLD structure (H).
All atomic lattice positions are extracted using 2D Gaussian fits for both the original and calculated reference lattice (E,F).
The difference between the fit positions at each lattice site yields the PLD displacements (I).
The complete data flow is summarized in (G).}
  \label{F:Method_Procedure}
\end{figure}

The algorithm used to produce periodic lattice displacement maps is summarized in Fig.~\ref{F:Method_Procedure}, and begins with a single high resolution STEM image (Fig.~\ref{F:Method_Procedure}c).
Let $I(\mathbf{r})$ be the input STEM image, where $\mathbf{r}\in\mathbb{R}^2$, let $\mathcal{R}$ be the set of all atomic column positions $\mathbf{R}$, and let $\delta(\mathbf{r})$ be the Dirac delta function.  We write the image $I(\mathbf{r})$ as
\begin{align}\label{E:image}
    I(\mathbf{r}) &= \sum_{\mathbf{R}\in\mathcal{R}} f\left(\mathbf{r}-\mathbf{R}\right) \nonumber\\
                  &= f\left(\mathbf{r}\right) \ast \sum_{\mathbf{R}\in\mathcal{R}} \delta\left(\mathbf{r} - \mathbf{R}\right)
\end{align}
where $\ast$ indicates a convolution and $f(\mathbf{r})$ is a form factor describing the STEM signal about each atomic site, incorporating the scattering cross section of high energy electrons with the projected potentials of the atomic columns, the finite point spread function of the electron beam, and channeling effects. 
For simplicity we consider the case of a single atomic species here, however, it is possible to include multiple form factors $f_i(\mathbf{r})$.

Atomic columns in STEM images are fit to two dimensional Gaussian functions and their positions extracted (Fig.~\ref{F:Method_Procedure}e).
This process can be considered a transformation which accepts an image $I(\mathbf{r})$ of the form in Eq.~(\ref{E:image}) and outputs the set of all atomic positions in the image, $\mathcal{R}$.
That is
\begin{equation*}
    \mathcal{G}\left(I(\mathbf{r})\right) = \mathcal{G}\left(f\left(\mathbf{r}\right) \ast \sum_{\mathbf{R}\in\mathcal{R}} \delta\left(\mathbf{r} - \mathbf{R}\right)\right) \equiv \mathcal{R}
\end{equation*}

Above, the set $\mathcal{R}$ is generic; let $\mathcal{R}^{\{0\}}$ be the set of all lattice points in an unmodulated lattice, which we here take for simplicity to be a Bravais lattice in two dimensions, $\mathcal{R}^{\{0\}}\equiv\{\mathbf{R}^{\{0\}}_{ij} = i\mathbf{a}_1 + j\mathbf{a}_2 \mid i,j \in \mathbb{Z}\}$.
For a lattice with a single sinusoidal modulation given by a modulation wavevector $\mathbf{q}_1$, we then write the set of all lattice points as $\mathcal{R}^{\{\mathbf{q}_1\}}$, for a lattice with two coexisting modulations $\mathbf{q}_1$ and $\mathbf{q}_2$ we write $\mathcal{R}^{\{\mathbf{q}_1, \mathbf{q}_2\}}$, and for a general set of modulation wavevectors $\mathcal{Q} \equiv \{\mathbf{q}_i \mid i \in 1 \ldots N\}$, we write $\mathcal{R}^{\mathcal{Q}}$ to indicate the set of all lattice points in the lattice modulated by all $\mathbf{q} \in \mathcal{Q}$.  Then $\mathcal{R}^{\mathcal{Q}} \equiv \{\mathbf{R}^\mathcal{Q}_{ij} \mid i,j \in \mathbb{Z}\}$, where modulated lattice sites may now be written in terms of the unmodulated lattice sites as
\begin{align}
    \mathbf{R}^{\mathbf{q}}_{ij} &= \mathbf{R}^{\{0\}}_{ij} + \mathbf{A}\sin\left(\mathbf{q}\cdot\mathbf{R}^{\{0\}}_{ij} + \phi\right)\nonumber\\
    \mathbf{R}^{\mathcal{Q}}_{ij} &= \mathbf{R}^{\{0\}}_{ij} + \sum_{\mathbf{q}\in\mathcal{Q}} \mathbf{A_{\mathbf{q}}}\sin\left(\mathbf{q}\cdot\mathbf{R}^{\{0\}}_{ij} + \phi_{\mathbf{q}}\right)\label{E:lattice_sites}
\end{align}
Here we focus on the case of sinusoidal modulations, but periodic modulations with more general waveforms are implicitly included by allowing $\mathcal{Q}$ to include higher order Fourier components.

The displacement of atomic column $(i,j)$ for a lattice with a single modulation vector $\mathbf{q}_1$ can then be written as $\mathbf{\Delta}^{\mathbf{q}_1}_{ij} = \mathbf{R}_{ij}^{\mathbf{q}_1} - \mathbf{R}_{ij}^{\{0\}}$.  
More generally, for a lattice with multiple coexisting modulations $\mathcal{Q}$, the displacement of each atomic column resulting solely from modulation vector $\mathbf{q}_p$ is
\begin{equation}\label{E:displacements}
    \mathbf{\Delta}^{\mathbf{q}_p}_{ij} = \mathbf{R}_{ij}^{\mathcal{Q}} - \mathbf{R}_{ij}^{\mathcal{Q}\backslash\mathbf{q}_p}
\end{equation}
where $\mathcal{Q}\backslash\mathbf{q}_p$ indicates the set $\mathcal{Q}$ with element $\mathbf{q}_p$ removed.
Thus, Eq.~(\ref{E:displacements}) simply defines the displacement at each atomic column due to a single modulation as the difference between atomic positions with and without that modulation present.

Once the lattice positions $\mathbf{R}^\mathcal{Q}_{ij}$ have been extracted by fitting Gaussians to each atomic site from an experimental image $I(\mathbf{r}) = I^{\mathcal{Q}}(\mathbf{r})$, via $\mathcal{G}\left(I^{\mathcal{Q}}(\mathbf{r})\right) = \mathcal{R}^{\mathcal{Q}}$, all that remains is to find the positions of a reference lattice in which the single modulation vector of interest $\mathbf{q}_p$ has been removed, $\mathcal{R}^{\mathcal{Q}\backslash\mathbf{q}_p}$.
This may be obtained by fitting the atomic columns of an image $I^{\mathcal{Q}\backslash\mathbf{q}_p}(\mathbf{r})$ in which $\mathbf{q}_p$ has been removed (Fig.~\ref{F:Method_Procedure}d), because $\mathcal{G}\left(I^{\mathcal{Q}\backslash\mathbf{q}_p}(\mathbf{r})\right) = \mathcal{R}^{\mathcal{Q}\backslash\mathbf{q}_p}$ (Fig.~\ref{F:Method_Procedure}f).
The task is therefore to obtain image $I^{\mathcal{Q}\backslash\mathbf{q}_p}(\mathbf{r})$ from an experimental image $I^{\mathcal{Q}}(\mathbf{r})$. 

Removing the contribution of a single modulation is most naturally approached in Fourier space.
Let $\mathcal{F}I(\mathbf{k}) = \int I(\mathbf{r})e^{-i\mathbf{k}\cdot\mathbf{r}}d\mathbf{r}$ be the Fourier transform of an image $I(\mathbf{r})$.
For an image $I^\mathcal{Q}(\mathbf{r})$ defined according to Eqs.~(\ref{E:image},\ref{E:lattice_sites}), the Fourier transform may be written as
\begin{align}\label{E:modulated_FT}
    \mathcal{F}I^{\mathcal{Q}}(\mathbf{k}) =
        & \mathcal{F}f(\mathbf{k}) \nonumber \\
        & \cdot \sum_{\alpha_{\mathbf{q}_1}=-\infty}^{\infty} \cdots \sum_{\alpha_{\mathbf{q}_N}=-\infty}^{\infty} \sum_{\mathbf{b}\in\mathcal{B}^{\{0\}}}\delta\left(\mathbf{k}-\left(\mathbf{b}-\sum_{\mathbf{q}\in\mathcal{Q}}\alpha_{\mathbf{q}}\mathbf{q}\right)\right) \nonumber \\
        & \quad\qquad\qquad\qquad\qquad\qquad\qquad\qquad\qquad \cdot \prod_{\mathbf{q}\in\mathcal{Q}} \mathcal{M}_{\alpha_{\mathbf{q}},\mathbf{q}}(\mathbf{k}) 
\end{align}
where $\mathcal{B}^{\{0\}}$ is the reciprocal lattice of $\mathcal{R}^{\{0\}}$, and 
\begin{equation}\label{E:k_dot_A_factor}
    \mathcal{M}_{\alpha_{\mathbf{q}},\mathbf{q}}(\mathbf{k}) \equiv J_{\alpha_{\mathbf{q}}}\left(\mathbf{k}\cdot\mathbf{A}_{\mathbf{q}}\right) \exp\left[i\alpha_{\mathbf{q}}\phi_{\mathbf{q}}\right] 
\end{equation}
where $J_\alpha(z)$ is the $\alpha$'th Bessel function of the first kind.
Derivation of Eq.~\ref{E:modulated_FT} is found at the end of this section, and related calculations are found in \cite{Wilson1975,hovden2016atomic}.
Here, the important observation is that for each of the Bragg peaks on the reciprocal lattice sites, $\delta\left(\mathbf{k}-\mathbf{b}\right)$ for $\mathbf{b}\in\mathcal{B}^{\{0\}}$, there are an additional set of satellite peaks offset from the Bragg peak by the linear combinations of the modulation vectors $\mathbf{q}\in\mathcal{Q}$, which encode the PLD.
The PLD is thus effectively decoupled from the underlying lattice in Fourier space.

In the experimental BSCMO STEM data here, only first order peaks for the two modulation vectors $\mathbf{q}_1$, $\mathbf{q}_2$ are observed, due to the damping of the higher order harmonics according to $\lvert\mathcal{M}_{\alpha}(\mathbf{k})\rvert \approx \frac{1}{\alpha}\left(\frac{1}{2}\mathbf{k}\cdot\mathbf{A}_\mathbf{q}\right)^\alpha$, where we've used the fact that the argument $\mathbf{k}\cdot\mathbf{A}_\mathbf{q} \ll 1$ (here, $\lvert\mathbf{A}_\mathbf{q}\rvert\approx 10$ pm and $\mathbf{k}\leq (1\text{\AA})^{-1}$, thus $\mathbf{k}\cdot\mathbf{A}_\mathbf{q} \leq 0.1$).
Including only the experimentally observable peaks reduces Eq.~\ref{E:modulated_FT} to
\begin{align}\label{E:simplified_FT}
    \mathcal{F}I^{\mathbf{q}_1,\mathbf{q}_2}(\mathbf{k}) =
        \mathcal{F}f(\mathbf{k}) \sum_{\mathbf{b}\in\mathcal{B}^{\{0\}}}
        c_0\delta\left(\mathbf{k}-\mathbf{b}\right) &+ c_{-,1}\delta\left(\mathbf{k}-\left(\mathbf{b}-\mathbf{q}_1\right)\right) + c_{+,1}\delta\left(\mathbf{k}-\left(\mathbf{b}+\mathbf{q}_1\right)\right)\nonumber\\
        &+ c_{-,2}\delta\left(\mathbf{k}-\left(\mathbf{b}-\mathbf{q}_2\right)\right) + c_{+,2}\delta\left(\mathbf{k}-\left(\mathbf{b}+\mathbf{q}_2\right)\right)
\end{align}
for complex constants $c$.

We then define a transformation $\mathcal{D}^{\mathbf{q}_p}$ which removes the contribution of modulation $\mathbf{q}_p$ from a Fourier transform.
By definition
\begin{equation*}
    \mathcal{D}^{\mathbf{q}_p}\left(\mathcal{F}I^{\mathcal{Q}}(\mathbf{k})\right) \equiv \mathcal{F}I^{\mathcal{Q}\backslash\mathbf{q}_p}(\mathbf{k})
\end{equation*}
Specifically
\begin{align*}			
	\mathcal{D}^{\mathbf{q}_1}\left(\mathcal{F}I^{\mathbf{q}_1,\mathbf{q}_2}(\mathbf{k})\right) = \mathcal{F}f(\mathbf{k}) \sum_{\mathbf{b}\in\mathcal{B}^{\{0\}}}
        &c_0\delta\left(\mathbf{k}-\mathbf{b}\right)\\ 
        &+ c_{-,2}\delta\left(\mathbf{k}-\left(\mathbf{b}-\mathbf{q}_2\right)\right) + c_{+,2}\delta\left(\mathbf{k}-\left(\mathbf{b}+\mathbf{q}_2\right)\right)\\
\mathcal{D}^{\mathbf{q}_2}\left(\mathcal{F}I^{\mathbf{q}_1,\mathbf{q}_2}(\mathbf{k})\right) = \mathcal{F}f(\mathbf{k}) \sum_{\mathbf{b}\in\mathcal{B}^{\{0\}}}
        &c_0\delta\left(\mathbf{k}-\mathbf{b}\right)\\ 
        &+ c_{-,1}\delta\left(\mathbf{k}-\left(\mathbf{b}-\mathbf{q}_1\right)\right) + c_{+,1}\delta\left(\mathbf{k}-\left(\mathbf{b}+\mathbf{q}_1\right)\right)
\end{align*}

Obtaining $\mathcal{D}^{\mathbf{q}_1}$ and $\mathcal{D}^{\mathbf{q}_2}$ thus requires carefully removing the relevant peaks from the experimental Fourier transform (Fig.~\ref{F:Method_Procedure}a).
Algorithmically, the positions of all detectable satellite peaks corresponding to a single modulation wavevector $\mathbf{q}$ of interest are extracted (Fig.~\ref{F:Method_Procedure}a, upper inset).
A mask radius is chosen, and the background level for each satellite peak is calculated by finding a 2D linear fit to the Fourier space amplitude in an annulus about that mask.
The amplitudes inside each masked region is then scaled down to this background level (Fig.~\ref{F:Method_Procedure}b, upper inset), while leaving the phase information unaltered (Fig.~\ref{F:Method_Procedure}a,b, lower insets), yielding $\mathcal{D}^{\mathbf{q}_p}$.  

An inverse Fourier transform is taken to obtain a $\mathbf{q}$--damped reference image,
\begin{equation*}    \mathcal{F}^{-1}\left(\mathcal{D}^{\mathbf{q}_p}\left(\mathcal{F}\left(I^{\mathcal{Q}}(\mathbf{r})\right)\right)\right) = I^{\mathcal{Q}\backslash\mathbf{q}_p}(\mathbf{r})
\end{equation*}
where $\mathcal{F}^{-1}$ is the inverse Fourier transform (Fig.~\ref{F:Method_Procedure}d).
The positions of all atomic sites in $I^{\mathcal{Q}\backslash\mathbf{q}_p}(\mathbf{r})$ are then extracted by fitting Gaussians to each site (Fig.~\ref{F:Method_Procedure}f), i.e.
\begin{equation*}    \mathcal{G}\left(\mathcal{F}^{-1}\left(\mathcal{D}^{\mathbf{q}_p}\left(\mathcal{F}\left(I^{\mathcal{Q}}(\mathbf{r})\right)\right)\right)\right) = \mathcal{R}^{\mathcal{Q}\backslash\mathbf{q}_p}
\end{equation*}

With both sets of atomic positions $\mathcal{R}^{\mathcal{Q}}$ and $\mathcal{R}^{\mathcal{Q}\backslash\mathbf{q}_p}$ in hand, $\Delta^{\mathbf{q}_p}_{i,j}(\mathbf{r})$ may then be directly calculate via Eq.~\ref{E:displacements} (Fig.~\ref{F:Method_Procedure}i).
A qualitative picture of the PLD structure may be obtained sans Gaussian fits by taking $I^\mathcal{Q}(\mathbf{r}) - I^{\mathcal{Q}\backslash\mathbf{q}_p}(\mathbf{r})$, shown in Fig.~\ref{F:Method_Procedure}h.
The complete data processing flow is summarized in Fig.~\ref{F:Method_Procedure}g.

As with any processing performed on raw data, in order to correctly interpret the results of this approach it is necessary to carefully understand precisely its limits, regime of validity, and any possible artifacts. 
The sections below discuss these points through a combination of simulation, experimental control datasets, and theoretical considerations.

\subsubsection*{Comparison with Direct, Real Space PLD Measurement}

\begin{figure}
  \includegraphics[width=5in]{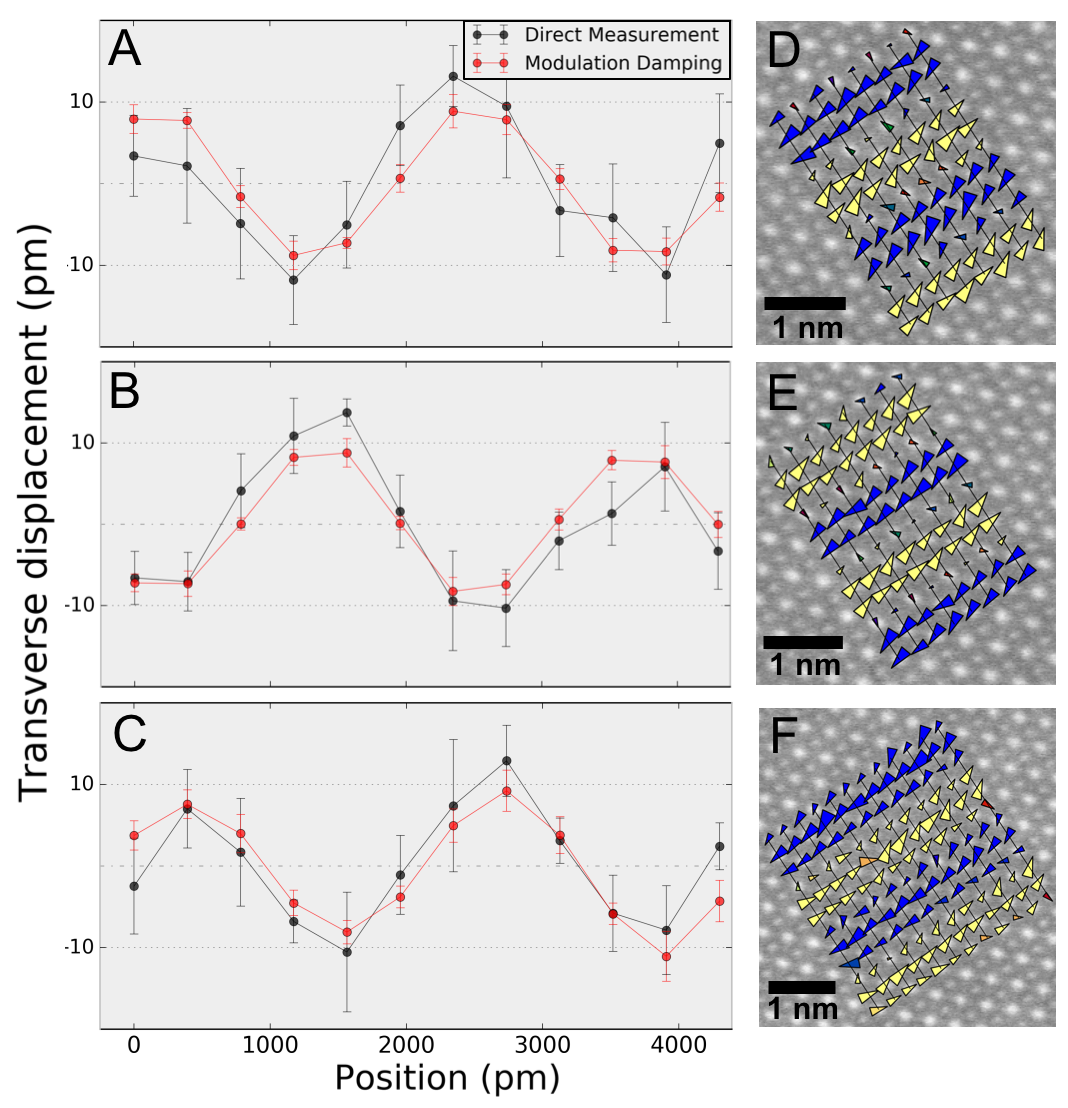}
  \caption{\textbf{Comparison to PLD measurement without Fourier damping.} 
Line profiles (A-C) of the transverse component of the displacements are calculated using both the modulation damping approach (\textit{red}) and by defining a local reference line in real space, without any Fourier space modifications (\textit{black}).
Circles and error bars represent the mean and standard deviations, respectively, of the transverse displacement measurements across a single row of 8-12 lattice sites in three well-ordered regions (D-F).
The two approaches yield consistent results.
We believe the larger error bars of the real-space approach primarily reflect that, unlike the modulation damping approach, this method does not account for image distortions, as well as the relative imprecision of using locally defined lines as reference positions.}
  \label{F:DirectMeasurement}
\end{figure}

In order to confirm that the measured PLDs are not an artifact of the Fourier space damping procedure, we calculated the displacement vectors directly from the data, unprocessed except for cross correlation, in real space.
In Fig.~\ref{F:DirectMeasurement}, the transverse component of the displacement vectors obtained using the modulation damping approach (red) are compared to the transverse distances of atomic centers from a line fit to the positions of all centers (black) along a column in the q-vector direction (lines, D-F) spanning two PLD wavelengths.
This real-space approach useful for confirming the validity of the method, however, it is only possible in well-ordered regions containing a single modulation.
Three different well-ordered regions of an experimental dataset, in which a single modulation dominates, are shown.  
Circles and error bars in the line profiles (Fig.~\ref{F:DirectMeasurement}A-C) represent the mean and standard deviations of the transverse displacements along a single row perpendicular to $\mathbf{q}$.

The results are in good agreement on average, clearly indicating that the modulation damping approach is indeed reconstructing a displacement field present in the data.
Notably, the real space approach has significantly larger error bars than the Fourier damping approach.
We attribute this to three factors.
First, the real space approach does not account for distortions resulting from the imaging process, limiting its accuracy.
Second, using a best-fit line as a reference position is a somewhat coarse approach, however, more systematic real-space methods such as a global coordinate rotation were untenable, likely due precisely to image distortion. 
Third, the reference lattice defined by the Fourier space method effectively represents a locally averaged reference structure, thus, sufficiently localized features in the displacement field may be smoothed out.
The following sections explore this last possibility in greater detail, and demonstrate that all but the very sharpest features in the PLD field are well described by the Fourier damping approach.

\subsubsection*{Method Limits, Regime of Validity, and Interpretation}

For a perfect lattice modulated by a perfect sinusoidal displacement field, both the Bragg and satellite peaks are delta functions.
Local disorder in the PLD field causes the satellite peaks to deviate from perfect impulses.
The size scale of the local features in the PLD structure relate to the degree of blurring observed in the satellite peaks, therefore the size of the damping mask used (Fig.~\ref{F:Method_Procedure}a,b \textit{upper insets}) determines the sharpness of the PLD features the method is able to reconstruct with good fidelity.
To capture the highest frequency variations in the PLD field possible, the largest mask size which does not interfere with other Fourier space features should be chosen.
In this work we used masks that extended halfway to the nearest Bragg peaks.
The mask radius is therefore approximately $\lvert\frac{1}{2}\mathbf{q}\rvert$, and we expect to correctly capture any PLD disorder features of size $\gtrsim 2\lambda_\text{PLD}$.
For PLD disorder of size scales $\lesssim 2\lambda_\text{PLD}$, our reconstructed displacements may deviate from the true displacement magnitudes somewhat. 
The simulations discussed below demonstrate that this effect is only appreciable at atomically sharp disorder in the PLD field, and we believe our reconstructed displacements are correct everywhere, with the possible exception of the atomic sites located precisely at topological defect cores.

\begin{figure}
  \includegraphics[width=6in]{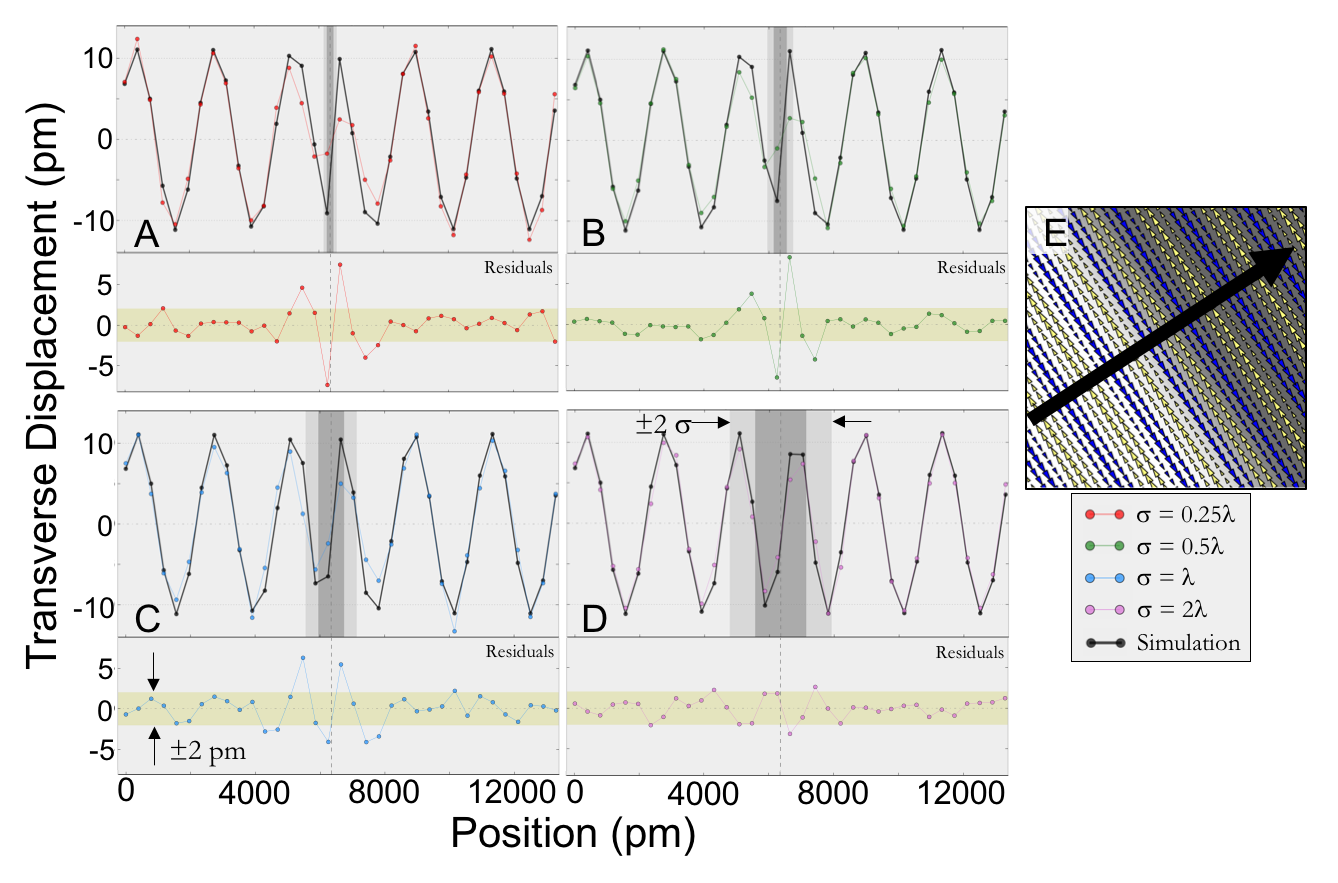}
  \caption{\textbf{Effect of abrupt PLD features.} 
Simulated data with antiphase domains in the PLD phase were generated with varying domain boundary width, by blurring a step function phase field with Gaussian kernels given by $\sigma=0.25\lambda$, $\sigma=0.5\lambda$, $\sigma=\lambda$, and $\sigma=2\lambda$ (A-D).
Line profiles (E) of the transverse displacement components are plotted for the simulated (\textit{black}) and calculated (\textit{color}) PLDs (A-D, \textit{above}), along with their corresponding residuals (A-D, \textit{below}).
Dark/light gray boxes in A-D indicate $\pm\sigma / \pm2\sigma$ about the domain boundary.
Dark/light regions in E indicate PLD phase values of 0/$\pi$, with the $\sigma=2\lambda$ case shown here.
The PLDs are accurately captured everywhere for $\sigma=2\lambda$.
The remaining cases all capture the PLDs accurately far from the domain boundaries.
For $\sigma=\lambda$ the phase jump is correctly captured qualitatively, but incorrectly damps the displacement amplitudes near the boundary.
For $\sigma=0.25\lambda$ and $\sigma=0.5\lambda$ the method fails at the atomic sites on the boundary.}
  \label{F:VaryingBlur}
\end{figure}

To carefully understand these effects and ensure correct interpretation of our results, we simulated data with a single sinusoidal modulation, and an antiphase domain boundary separating regions in which the PLD phase shifts by $\pi$.
Four datasets were simulated, varying the abruptness of the antiphase domain boundary by first generating a step function domain boundary and then blurring with a Gaussian kernel with four different values of $\sigma$.
Fig.~\ref{F:VaryingBlur} A-D show the results for $\sigma=0.25\lambda_\text{PLD}$, $\sigma=0.5\lambda_\text{PLD}$, $\sigma=\lambda_\text{PLD}$, and $\sigma=2\lambda_\text{PLD}$, respectively.
Each shows the simulated and calculated transverse displacement components along the line profile shown in Fig.~\ref{F:VaryingBlur}E, along with their residuals.
In every case, the residuals fall inside our estimated $\sim\pm2$ pm precision within a few lattice spacings of the interface.
Further, in every case the reconstructed displacements accurately capture the qualitative structure of the antiphase boundary everywhere except for at the lattice sites precisely at the center of the interface.
Quantitatively, the reconstructed displacements accurately capture the simulated data everywhere for the $\sigma=2\lambda_\text{PLD}$ case.
For smaller $\sigma$, the residuals are somewhat larger within $\sim\lambda_\text{PLD}$ about the interface, as expected.
For very sharp interfaces ($\sigma=0.25\lambda_\text{PLD}$ and $\sigma=0.5\lambda_\text{PLD}$) the reconstructed displacements are incorrect at the center of the interface, where the atomically sharp discontinuity in the simulated displacement field is averaged out in the reconstruction to yield incorrectly small displacements.

We conclude that for local PLD disorder of size $\gtrsim2\lambda_\text{PLD}$, our approach is valid everywhere.
At smaller features, our approach correctly captures qualitative structure but tends to underestimate displacement amplitudes near the feature center.
The approach fails entirely only at atomically abrupt features in the PLD field.
We contend that this approach therefore reasonably reflects the PLD structure everywhere in the data presented, with the possible exception of the topological defect cores.
We believe the reconstructed displacements about defect cores are likely correct, because of smoothing and amplitude damping of the total displacement vectors near phase singularities, however, we cannot discount the possibility that we have averaged over a sharp discontinuity at a defect core.
In this scenario, the reconstructed displacements would only be incorrect at the sites directly adjacent to the core, and the amplitudes at these sites may be considered a lower bound on their true displacement amplitudes.
Note that we cannot experimentally discount atomically sharp disorder elsewhere in the PLD field, however, we believe such features to be unlikely on energetic grounds.

\subsubsection*{Effect of Fourier Mask Size}
\begin{figure}
  \includegraphics[width=6in]{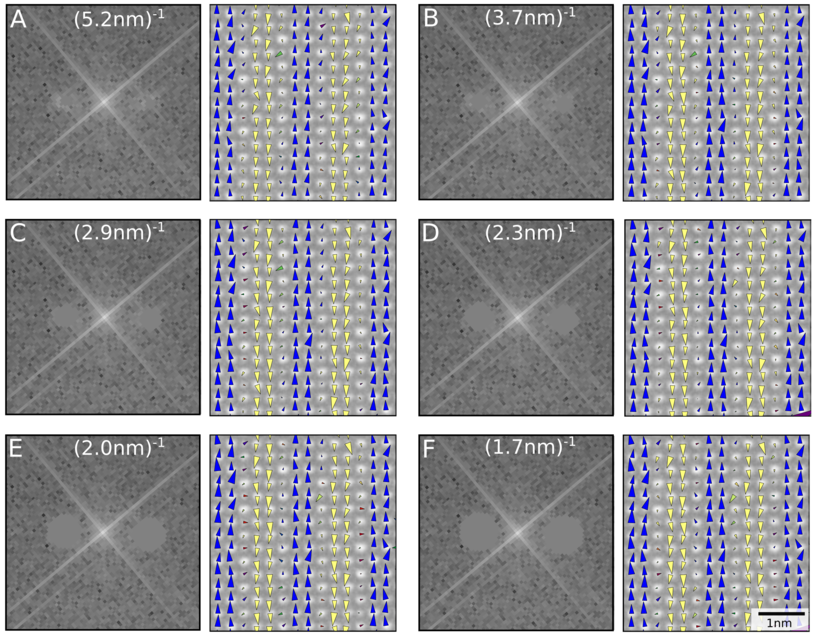}
  \caption{\textbf{Effect of varying mask size on real data.} 
(\textbf{A-F}) Typical Fourier peaks with reciprocal space mask diameter varying from $(5.2\text{ nm})^{-1}$ to $(1.7\text{ nm})^{-1}$ and their respective PLD maps. 
When the mask size is too small (A), the amplitude of the PLD is diminished since the mask does not reflect the total intensity of the satellite peak.
Once the peak is fully captured by the mask, the mapping is insensitive to increasing mask size.
}
\label{F:VaryingMask_real}
\end{figure}

\begin{figure}
  \includegraphics[width=6in]{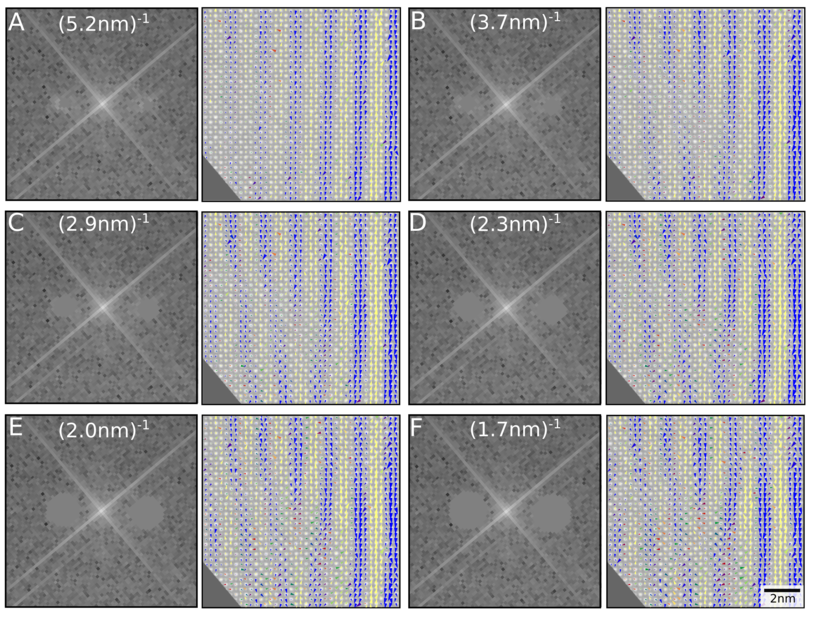}
  \caption{\textbf{Effect of varying mask size on mapping of a topological defect.} 
(\textbf{A-F}) Typical Fourier peaks with reciprocal space mask diameter varying from $(5.2\text{ nm})^{-1}$ to $(1.7\text{ nm})^{-1}$ and the respective PLD maps of a topological defect. 
The topological defect is missing in A because the mask does not fully capture the satellite peak. 
For mask sizes that fully capture by the satellite peaks, the mapping is insensitive to increasing mask size.
}
\label{F:VaryingMask_real_topo}
\end{figure}

We tested the effect of varying the mask size on experimental data, in both a well ordered region and a disordered region containing a topological defect, shown in Figs.~\ref{F:VaryingMask_real} and Fig.~\ref{F:VaryingMask_real_topo}, respectively.
In both cases, a very small mask does not damp the full intensity of the satellite peak.
In ordered regions, this leads to artificially reduced PLD amplitude relative to the other PLD maps of the same data (Figs.~\ref{F:VaryingMask_real}A and \ref{F:VaryingMask_real_topo}A).
A very large mask begins to introduce greater noise, observable at the modulation minima, and likely resulting from beginning to damp some of the Fourier space intensity in the tails of the nearby Bragg peak.
In the intermediate range of mask sizes which cover the entire satellite peak but remain far from the Bragg peak, the reconstructed PLD is insensitive to mask size variation.  
We additionally tested the effect of introducing noise into the Fourier space damping level comparable to the noise observed nearby in Fourier space, with negligible effect.

\begin{figure}
  \includegraphics[width=6in]{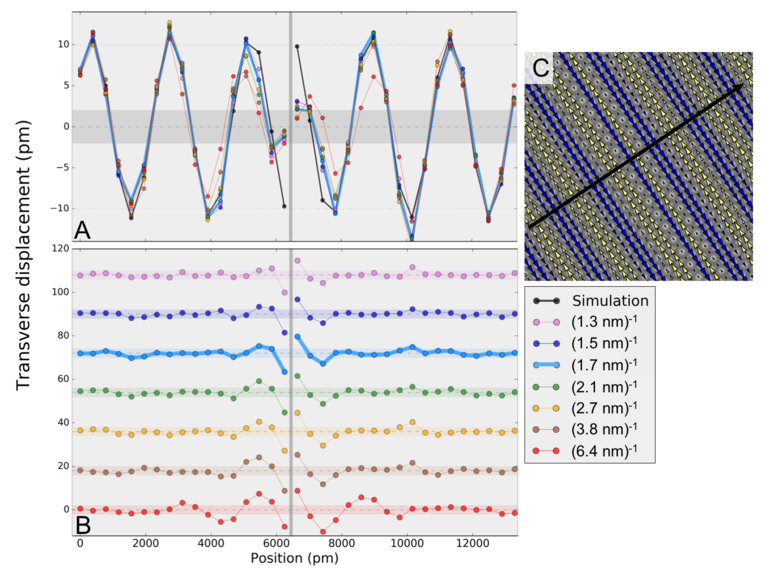}
  \caption{\textbf{Effect of varying mask size on simulated data.} 
Simulated data with a Heavyside function antiphase domain boundary in the PLD phase was generated, and the PLD reconstructed using 7 different mask sizes.
Line profiles of the transverse displacement components are plotted for the simulated (\textit{black}) and calculated (\textit{color}) PLDs (A), along with their corresponding residuals (B), for the sites indicated in the PLD map (C).
The horizontal bars in A (\textit{gray}) and B (\textit{colors}) indicate $\pm2$ pm.
All mask sizes accurately capture the displacements far from the interface, and fail to capture the true displacements of the two sites at the atomically sharp antiphase interface. 
Very small Fourier space masks (\textit{red, brown}) result in artifacts in the residuals several lattice spacings or more from the interface.
For large Fourier space masks (\textit{light blue, dark blue, purple}) the residuals are on the order of the $\pm2$ pm error bars 2 lattice spacings from the interface.
The typical mask size used on experimental data is bolded (\textit{light blue}, $(1.7\text{ nm})^{-1}$.}
  \label{F:VaryingMask_sim}
\end{figure}

In order to further understand the effect of the Fourier mask size, we varied the mask size used to reconstruct a simulated dataset containing an antiphase domain boundary in the PLD phase.
To test how our PLD reconstruction approach behaves in the worst case scenario, in this simulated data the $0$ to $\pi$ transition in the PLD phase is given by a perfect step function.
Fig.~\ref{F:VaryingMask_sim} shows the results for 7 mask sizes, corresponding to real space diameters of 1.3 nm, 1.5 nm, 1.7 nm, 2.1 nm, 2.7 nm, 3.8 nm, and 6.4 nm.
In Fig.~\ref{F:VaryingMask_sim} A, the transverse components of the simulated displacements (\textit{black}) agree well with their reconstructed counterparts far from the boundary.
The single atomic sites directly adjacent to either side of the boundary are incorrect in all cases, as expected for an atomically sharp feature and discussed in the previous section.
The fidelity of the reconstruction in the intermediate region is evident in the residuals, Fig.~\ref{F:VaryingMask_sim} B.
The vertical scale is indicated by the colored horizontal bars, which correspond to the gray horizontal bar in Fig.~\ref{F:VaryingMask_sim} A and represent the $\pm2$ pm error bars.
The smallest Fourier space mask (\textit{red}, 6.4 nm) displays damped periodic ringing in the residuals reminiscent of a sinc function.
This artifact is gradually reduced as the mask size is increased (and its corresponding real space distance is decreased), with residuals for the largest masks falling inside the $\pm$2 pm error bars within 2-3 lattice spacings of the interface.
The typical mask size used on experimental data, corresponding to 1.7 nm in real space, is highlighted (\textit{bold, light blue}), and captures the true displacements well everywhere except at the atomic sites directly adjacent to the boundary.
We re-iterate that while a simulated step-function $\pi$ phase slip is useful to evaluate the effectiveness of our approach, energetically we would not expect such high frequency features in experimental PLDs, with the exception of topological defect cores.

\subsubsection*{Method Comparison with SrTiO$_{3}$}
\begin{figure}
  \includegraphics[width=6in]{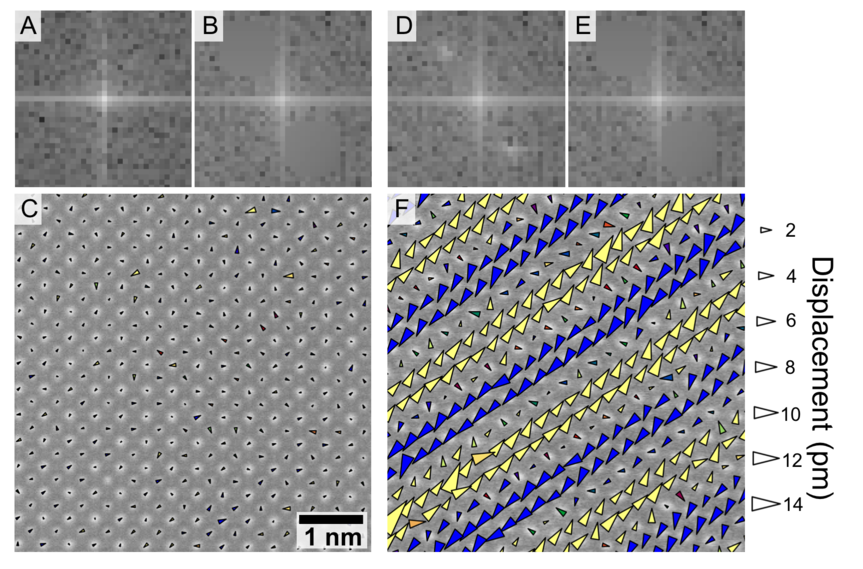}
  \caption{\textbf{PLD mapping in STO vs BSCMO} 
The Fourier transform of SrTiO$_{3}$ (STO) data does not contain satellite peaks about the Bragg peaks (A), while the Fourier transform of BSCMO data does (D).
The damping procedure was performed on both STO data (B) and BSCMO data (E) at comparable positions in Fourier space.
The resulting PLD map for STO data shows displacement vectors with a mean magnitude of 0.390 pm and no clear periodic structure (C), in stark contrast to the strong PLD structure observed for BSCMO PLD map (F).
The image and displacement vector size scales are identical for (C) and (F).}
  \label{F:STO_v_BSCMO}
\end{figure}

To ensure our Fourier damping approaching was not introducing artificial periodic structure or other artifacts in the lattice displacements, we performed our method on STEM data of cubic SrTiO$_{3}$ (STO).
The results are shown with identical analysis of comparable BSCMO data in Fig.~\ref{F:STO_v_BSCMO}.
A single Bragg peak from the Fourier transform of the STO data (Fig.~\ref{F:STO_v_BSCMO}A) shows no satellite peaks, while a single Bragg peak from the Fourier transform of the BSCMO data has two satellite peaks (Fig.~\ref{F:STO_v_BSCMO}D) corresponding to a single PLD modulation in this dataset.
After finding the $\mathbf{q}$-vector for the BSCMO data, an equivalent vector scaled to the STO reciprocal lattice was calculated.
The peak damping procedure was then performed for both the STO and BSCMO data (Fig.~\ref{F:STO_v_BSCMO}B,E), and PLD maps were generated (Fig.~\ref{F:STO_v_BSCMO}C,F).
The scalebar and displacement vector scales are identical for the two PLD maps.
The BSCMO data shows periodic stripes of $\sim$10 pm transverse displacements. 
In the STO data the mean displacement magnitude is 0.390 pm, 90\% of the displacement magnitudes are $\leq 0.83$ pm, 99\% of the displacements magnitudes are $\leq 2.08$ pm, and no clear periodicity is observed.

\subsubsection*{Atomic fit precision with STO}

\begin{figure}
  \includegraphics[width=6in]{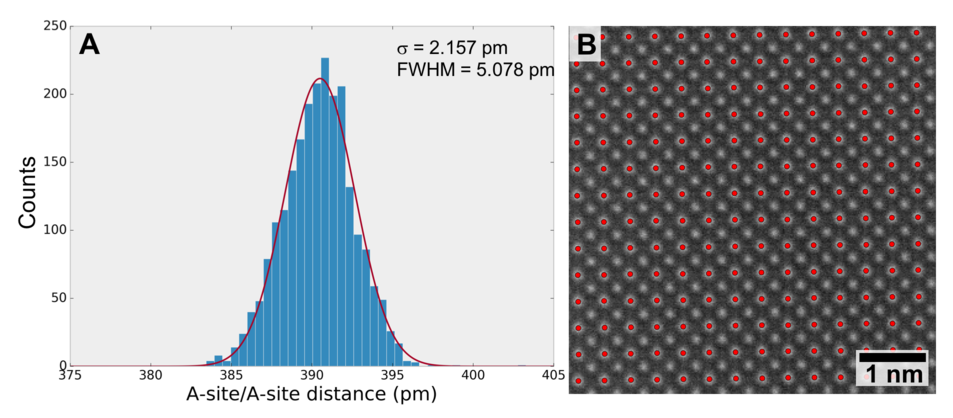}
  \caption{\textbf{Precision of atomic position fits} 
\textbf{(A)} A histogram of distances between neighboring Sr sites in STO data \textbf{(B)} is used to estimate the precision of the atomic positions extracted via 2D Gaussian fits.
A 1D Gaussian fit to the resulting histogram has $\sigma=2.157$ pm and FWHM$=5.078$ pm.
Identical analysis for the lower signal-to-noise ratio Ti sites yields $\sigma=2.415$ pm and FWHM$=5.687$ pm.}
  \label{F:fit_precision}
\end{figure}

In order to determine the precision of our atomic fitting, we fit the positions of all atomic sites in STO data, and calculated distances between neighboring atomic sites.
A histogram of the distances between the fit positions of nearest neighbor Sr atoms and the data used are shown in Fig.~\ref{F:fit_precision}A,B.
A Gaussian fit to the histogram (Fig.~\ref{F:fit_precision}A) has a standard deviation of $\sigma=2.157$ pm and a full width at half maximum of $5.1$ pm.
Identical analysis of the dimmer, lower signal-to-noise ratio Ti sites yields values of $\sigma=2.4$ pm and a FWHM of $5.7$ pm.
While our precision is comparable to that obtained by others, notably it is significantly worse than the 0.6 pm precision obtained in \cite{Yankovich2014}.
We attribute our lower precision to image acquisition which has been optimized for different purposes.
Here, we aimed to obtain high precision atomic fits, large fields of view, and optimal Fourier space sampling for satellite peak damping.

\subsubsection*{Derivation of Eq.~\ref{E:modulated_FT}}

Taking the Fourier transform of an image given by Eq.~\ref{E:image} with modulated lattice sites defined in Eq.~\ref{E:lattice_sites}, we find
\begin{align*}
    \mathcal{F}I^{\mathcal{Q}}(\mathbf{k}) &= \mathcal{F}\left(f(\mathbf{r})\ast\sum_{\mathbf{R}\in\mathcal{R}^{\mathcal{Q}}}\delta\left(\mathbf{r}-\mathbf{R}\right)\right) \\
        &=\mathcal{F}\left(f(\mathbf{r})\right)\mathcal{F}\left(\sum_{\mathbf{R}\in\mathcal{R}^{\mathcal{Q}}}\delta\left(\mathbf{r}-\mathbf{R}\right)\right) \\
        &=\mathcal{F}f(\mathbf{k})\sum_{\mathbf{R}\in\mathcal{R}^{\mathcal{Q}}}\left(\mathcal{F}\left(\delta\left(\mathbf{r}-\mathbf{R}\right)\right)\right) \\
        &=\mathcal{F}f(\mathbf{k})\sum_{\mathbf{R}\in\mathcal{R}^{\mathcal{Q}}}\left( e^{i\mathbf{k}\cdot\mathbf{R}} \right)
\end{align*}

Using Eq.~(\ref{E:lattice_sites}),
\begin{align*}
    \sum_{\mathbf{R}\in\mathcal{R}^{\mathcal{Q}}}\left( e^{i\mathbf{k}\cdot\mathbf{R}} \right) 
        &= \sum_{\mathbf{R}\in\mathcal{R}^{\{0\}}} \exp\left[i\mathbf{k}\cdot\left(\mathbf{R} 
         + \sum_{\mathbf{q}\in\mathcal{Q}}\mathbf{A}_{\mathbf{q}}\sin\left(\mathbf{q}\cdot\mathbf{R}+\phi_{\mathbf{q}}\right)\right)\right] \\
        &= \sum_{\mathbf{R}\in\mathcal{R}^{\{0\}}} \exp\left[i\mathbf{k}\cdot\mathbf{R}\right] \exp\left[i\mathbf{k}\cdot\sum_{\mathbf{q}\in\mathcal{Q}}\mathbf{A}_{\mathbf{q}}\sin\left(\mathbf{q}\cdot\mathbf{R}+\phi_{\mathbf{q}}\right)\right]\\
        &= \sum_{\mathbf{R}\in\mathcal{R}^{\{0\}}} \exp\left[i\mathbf{k}\cdot\mathbf{R}\right] \prod_{\mathbf{q}\in\mathcal{Q}}\exp\left[i\mathbf{k}\cdot\left(\mathbf{A}_{\mathbf{q}}\sin\left(\mathbf{q}\cdot\mathbf{R}+\phi_{\mathbf{q}}\right)\right)\right]
\end{align*}
The Jacobi-Anger expansion may be written as
\begin{equation*}
    e^{iz\sin\theta} = \sum_{\alpha=-\infty}^{\infty}J_\alpha(z)e^{i\alpha\theta}
\end{equation*}
where $J_\alpha(z)$ is the $\alpha$'th Bessel function of the first kind.  Then
\begin{align*}
    \sum_{\mathbf{R}\in\mathcal{R}^{\mathcal{Q}}}\left( e^{i\mathbf{k}\cdot\mathbf{R}} \right) 
        &= \sum_{\mathbf{R}\in\mathcal{R}^{\{0\}}} \exp\left[i\mathbf{k}\cdot\mathbf{R}\right] \nonumber \\
        & \qquad \cdot\prod_{\mathbf{q}\in\mathcal{Q}} \sum_{\alpha=-\infty}^{\infty}J_\alpha\left(\mathbf{k}\cdot\mathbf{A}_{\mathbf{q}}\right) \exp\left[i\alpha\mathbf{q}\cdot\mathbf{R}\right]\exp\left[i\alpha\phi_{\mathbf{q}}\right]
\end{align*}
Expanding the product over the $N$ elements in $\mathcal{Q}$,
\begin{align*}
    \sum_{\mathbf{R}\in\mathcal{R}^{\mathcal{Q}}}\left( e^{i\mathbf{k}\cdot\mathbf{R}} \right) 
        &= \sum_{\mathbf{R}\in\mathcal{R}^{\{0\}}} \exp\left[i\mathbf{k}\cdot\mathbf{R}\right] \sum_{\alpha_{\mathbf{q}_1}=-\infty}^{\infty} \cdots \sum_{\alpha_{\mathbf{q}_N}=-\infty}^{\infty} \prod_{\mathbf{q}\in\mathcal{Q}} \bigg( J_{\alpha_{\mathbf{q}}}\left(\mathbf{k}\cdot\mathbf{A}_{\mathbf{q}}\right) \\
        & \qquad\qquad\qquad\qquad\qquad\qquad\qquad \exp\left[i\alpha_{\mathbf{q}}\mathbf{q}\cdot\mathbf{R}\right]\exp\left[i\alpha_{\mathbf{q}}\phi_{\mathbf{q}}\right] \bigg)\\
        &= \sum_{\alpha_{\mathbf{q}_1}=-\infty}^{\infty} \cdots \sum_{\alpha_{\mathbf{q}_N}=-\infty}^{\infty} \Bigg( \sum_{\mathbf{R}\in\mathcal{R}^{\{0\}}} \exp\left[i\mathbf{k}\cdot\mathbf{R}\right]  \exp\left[\sum_{\mathbf{q}\in\mathcal{Q}} i\alpha_{\mathbf{q}}\mathbf{q}\cdot\mathbf{R}\right] \\
        & \qquad \cdot \prod_{\mathbf{q}\in\mathcal{Q}} \bigg( J_{\alpha_{\mathbf{q}}}\left(\mathbf{k}\cdot\mathbf{A}_{\mathbf{q}}\right) \exp\left[i\alpha_{\mathbf{q}}\phi_{\mathbf{q}}    \right] \bigg) \Bigg) \\
        &= \sum_{\alpha_{\mathbf{q}_1}=-\infty}^{\infty} \cdots \sum_{\alpha_{\mathbf{q}_N}=-\infty}^{\infty} \Bigg( \sum_{\mathbf{R}\in\mathcal{R}^{\{0\}}} \exp\left[i\left(\mathbf{k}+\sum_{\mathbf{q}\in\mathcal{Q}}\alpha_{\mathbf{q}}\mathbf{q}\right)\cdot\mathbf{R}\right] \\
        & \qquad \cdot \prod_{\mathbf{q}\in\mathcal{Q}} \bigg( J_{\alpha_{\mathbf{q}}}\left(\mathbf{k}\cdot\mathbf{A}_{\mathbf{q}}\right) \exp\left[i\alpha_{\mathbf{q}}\phi_{\mathbf{q}}    \right] \bigg) \Bigg) \\
        &= \sum_{\alpha_{\mathbf{q}_1}=-\infty}^{\infty} \cdots \sum_{\alpha_{\mathbf{q}_N}=-\infty}^{\infty} \Bigg( \sum_{\mathbf{b}\in\mathcal{B}^{\{0\}}}\delta\left(\mathbf{k}-\left(\mathbf{b}-\sum_{\mathbf{q}\in\mathcal{Q}}\alpha_{\mathbf{q}}\mathbf{q}\right)\right) \\
        & \qquad \cdot \prod_{\mathbf{q}\in\mathcal{Q}} \bigg( J_{\alpha_{\mathbf{q}}}\left(\mathbf{k}\cdot\mathbf{A}_{\mathbf{q}}\right) \exp\left[i\alpha_{\mathbf{q}}\phi_{\mathbf{q}}    \right] \bigg) \Bigg)
\end{align*}
where in the last step we've used the fact that $\sum_{\mathbf{R}\in\mathcal{R}^{\{0\}}}\exp\left[i\mathbf{k}\cdot\mathbf{R}\right] = \sum_{\mathbf{b}\in\mathcal{B}^{\{0\}}}\delta\left(\mathbf{k}-\mathbf{b}\right)$, where $\mathcal{B}^{\{0\}}$ is the reciprocal lattice of Bravais lattice $\mathcal{R}^{\{0\}}$.

Defining
\begin{equation*}
    \mathcal{M}_{\alpha_{\mathbf{q}},\mathbf{q}}(\mathbf{k}) \equiv J_{\alpha_{\mathbf{q}}}\left(\mathbf{k}\cdot\mathbf{A}_{\mathbf{q}}\right) \exp\left[i\alpha_{\mathbf{q}}\phi_{\mathbf{q}}\right] 
\end{equation*}
yields Eq.~\ref{E:modulated_FT}.

\subsection*{Coarse Grained Phase Field Extraction}

We next describe the Fourier space approach to extract the coarse-grained phase field, $\phi(\mathbf{r})$, associated with the $\mathbf{q_{1}}$ modulation
\begin{equation*}
\mathbf{\Delta^{\mathbf{q}_1}} \sim \sin(\mathbf{q}_{1}\cdot\mathbf{r}+\phi(\mathbf{r}))
\end{equation*}
We interpret the phase field as deviations of the $\mathbf{q}_1$ modulation from perfect periodicity (i.e. where $\phi(\mathbf{r}) = \phi_{0} = \text{const}$). 
We first Fourier filter regions near a $\mathbf{q}_1$ superlattice peak, typically one near the 200 (002) Bragg peak, using a Gaussian filter with a width $\sigma = L^{-1}$ where $L$ is the coarsening length scale in real space.
We obtain a real space image where all periodicities in the image, except for the one associated with the $\mathbf{q}_1$ modulation, are removed.
Roughly, the filtered image may be described by 
\begin{equation*}
\tilde{I}(\mathbf{r}) \sim \sin(\mathbf{q}_1\cdot\mathbf{r}+\phi(\mathbf{r}))
\end{equation*}
In order to extract $\phi(\mathbf{r})$, we use the phase lock-in technique described in \cite{Lawler2010}, where we generate two reference signals $\sin(\mathbf{q}_1\cdot\mathbf{r})$ and $\cos(\mathbf{q}_1\cdot\mathbf{r})$ with perfect $\mathbf{q}_1$ periodicity and multiply them by $\tilde{I}(\mathbf{r})$ to get $X(\mathbf{r})$ and $Y(\mathbf{r})$ where

\begin{equation*}
\begin{aligned}
\begin{cases}
X(\mathbf{r}) = \sin(\mathbf{q}_1\cdot\mathbf{r}) \sin(\mathbf{q}_1\cdot\mathbf{r}+\phi(\mathbf{r})) \\
Y(\mathbf{r}) = \cos(\mathbf{q}_1\cdot\mathbf{r}) \sin(\mathbf{q}_1\cdot\mathbf{r}+\phi(\mathbf{r}))
\end{cases}
\end{aligned}
\end{equation*}

\begin{equation*}
\begin{aligned}
\begin{cases}
X(\mathbf{r}) = \frac{1}{2}\left(\cos\phi(\mathbf{r}) - \cos(2\mathbf{q}_1\cdot\mathbf{r}+\phi(\mathbf{r}))\right) \\
Y(\mathbf{r}) = \frac{1}{2}\left(\sin\phi(\mathbf{r}) + \sin(2\mathbf{q}_1\cdot\mathbf{r}+\phi(\mathbf{r}))\right)
\end{cases}
\end{aligned}
\end{equation*}
We subsequently low pass filter $X(\mathbf{r})$ and $Y(\mathbf{r})$ to get rid of the second high frequency terms obtaining: 
\begin{equation*}
\begin{aligned}
\begin{cases}
\tilde{X}(\mathbf{r}) \approx \cos\phi(\mathbf{r})  \\
\tilde{Y}(\mathbf{r}) \approx \sin\phi(\mathbf{r}) 
\end{cases}
\end{aligned}
\end{equation*}
The coarse-grained phase is thus given by 
\begin{equation*}
\phi(\mathbf{r}) = \arctan[\tilde{Y}(\mathbf{r})/ \tilde{X}(\mathbf{r})]
\end{equation*}

The coarsening length must be chosen judiciously in order to simultaneously optimize the resolution and signal to noise ratio of the resulting coarse grained phase field.

\subsection*{Transverse vs. Longitudinal PLDs in Fourier Space}
PLDs have several Fourier space features that are distinct from the Fourier space structure of similar phenomena, including charge density waves and superlattices of atomic species.
Distinguishing transverse from longitudinal PLDs is readily accomplished in Fourier space by observing the intensity pattern of the satellite peaks with varying $\mathbf{k}$.
For simplicity, consider Eq.~\ref{E:simplified_FT} describing the experimentally observed peaks.
Using Eq.~\ref{E:k_dot_A_factor}, the factors damping each satellite peak are given here by
\begin{align*}
	c_{{\pm},i} &= J_1\left(\mathbf{k}\cdot\mathbf{A}_{\mathbf{q}_i}\right)\exp\left(\pm i\phi_{\mathbf{q}_i}\right)\\
    \lvert c_{{\pm},i}\rvert &\approx \mathbf{k}\cdot\mathbf{A}_{\mathbf{q}_i}
\end{align*}

\begin{figure}
  \includegraphics[width=6in]{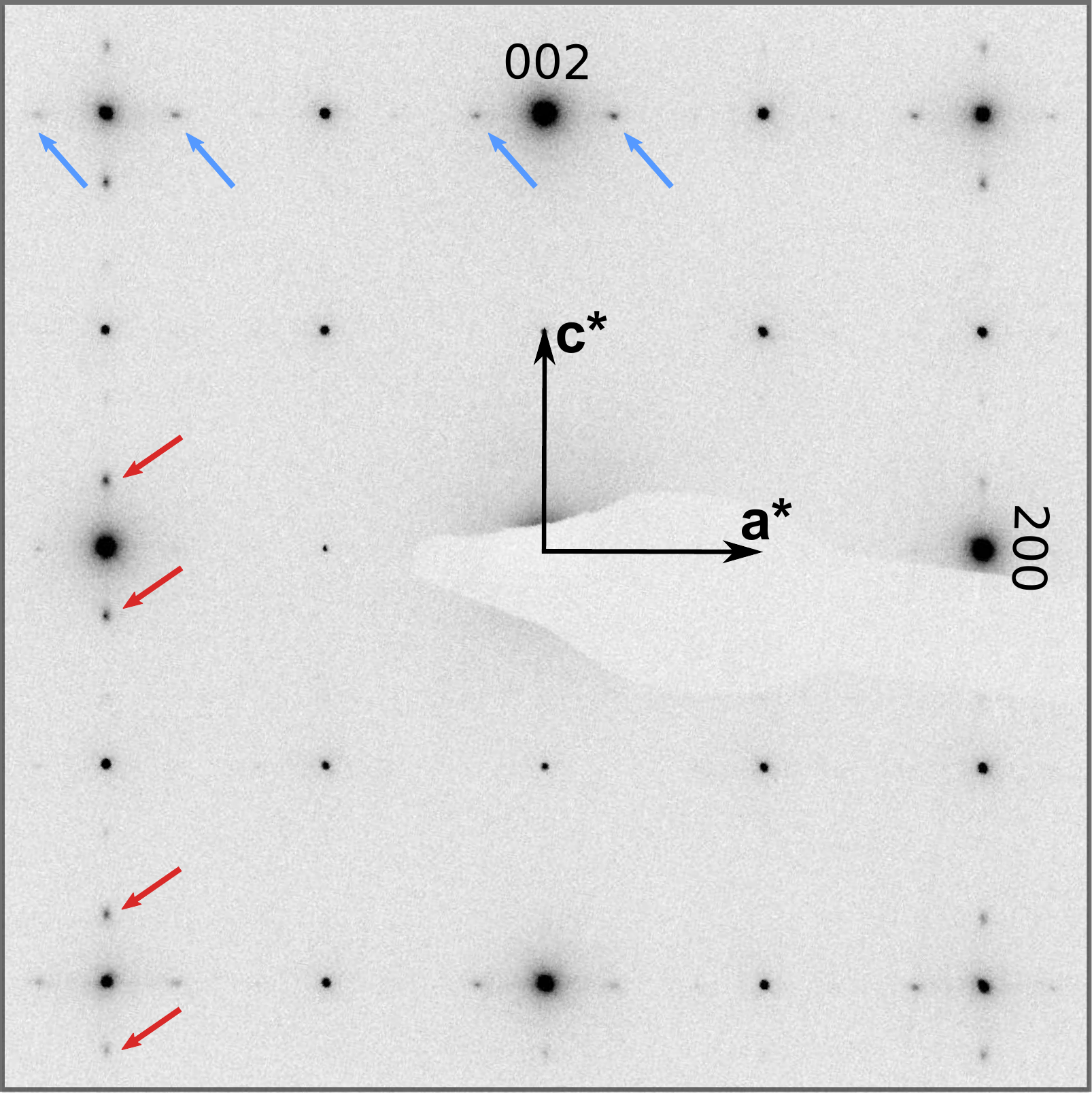}
  \caption{\textbf{Electron diffraction of BSCMO.} 
Electron diffraction pattern over 1$\mu m$ area indexed in the $Pnma$ space group.
Satellite peaks corresponding to modulations along orthogonal directions are marked by blue and red arrows.
Transverse, displacive lattice modulations are indicated.}
\label{F:Diffraction}
\end{figure}

\begin{figure}
  \includegraphics[width=6in]{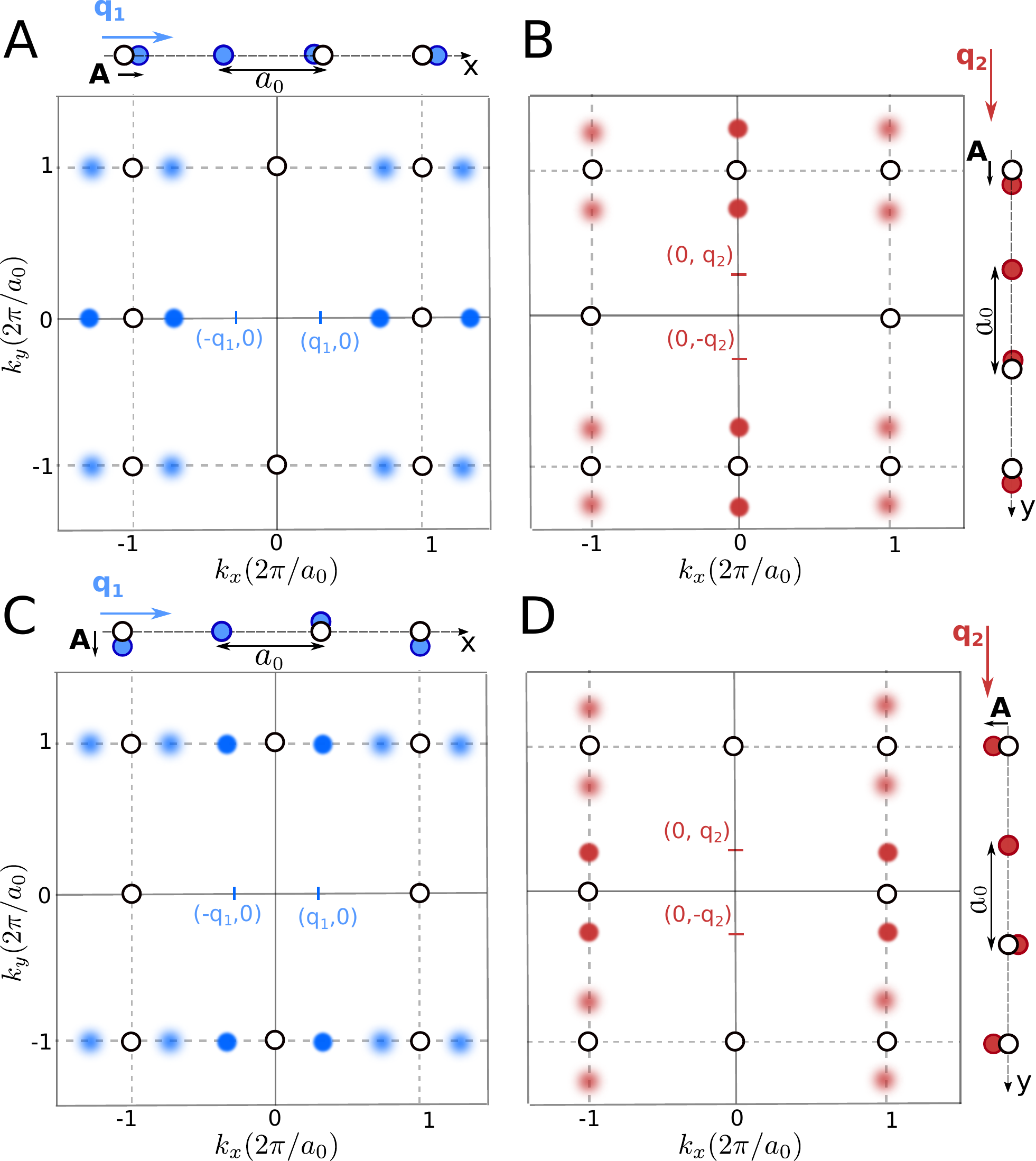}
  \caption{\textbf{Longitudinal vs. transverse PLDs in Fourier space.} 
(\textbf{A,B}) Reciprocal space structure of a square lattice modulated by a longitudinal, displacive modulation along $x$ and $y$, respectively.
(\textbf{C,D}) Reciprocal space structure of a square lattice modulated by a transverse, displacive modulation along $x$ and $y$, respectively.
STEM Fourier transforms and diffraction of BSCMO indicate transverse PLDs.}
  \label{F:TransLong}
\end{figure}

For some modulation wavevector $\mathbf{q}$, consider the satellite peaks about a Bragg peak $\mathbf{b}_{\parallel}$ parallel to the modulation $\mathbf{q} \parallel \mathbf{b}_\parallel$.  
At the satellite positions $\mathbf{k}\approx\mathbf{b}_\parallel$ the damping factor is then $\lvert c_{{\pm},\parallel}\rvert \approx \mathbf{b}_\parallel\cdot\mathbf{A}_{\mathbf{q}}$.
Thus $\lvert c_{{\pm},\parallel}\rvert$ is maximal for a longitudinal PLD where $\mathbf{q}\parallel\mathbf{A}_\mathbf{q}$, while $\lvert c_{{\pm},\parallel}\rvert \approx 0$ and the satellite peaks vanish for a transverse PLD where $\mathbf{q}\perp\mathbf{A}_\mathbf{q}$.
In contrast, consider the satellite peaks about a Bragg peak $\mathbf{b}_{\perp}$, perpendicular to the modulation vector $\mathbf{q} \perp \mathbf{b}_\parallel$.  
Now $\lvert c_{{\pm},\perp}\rvert$ is maximal for a transverse PLD where $\mathbf{q}\perp\mathbf{A}_\mathbf{q}$, while $\lvert c_{{\pm},\perp}\rvert \approx 0$ and the satellite peaks vanish for a longitudinal PLD where $\mathbf{q}\parallel\mathbf{A}_\mathbf{q}$.
These cases are illustrated schematically in Fig.~\ref{F:TransLong}.
Both the STEM Fourier transforms and diffraction patterns (Fig.~\ref{F:Diffraction}) of BSCMO clearly indicate transverse PLDs in BSCMO.

\newpage
\section*{Supplemental Figures}

\begin{figure}[h]
  \includegraphics[width=5.2in]{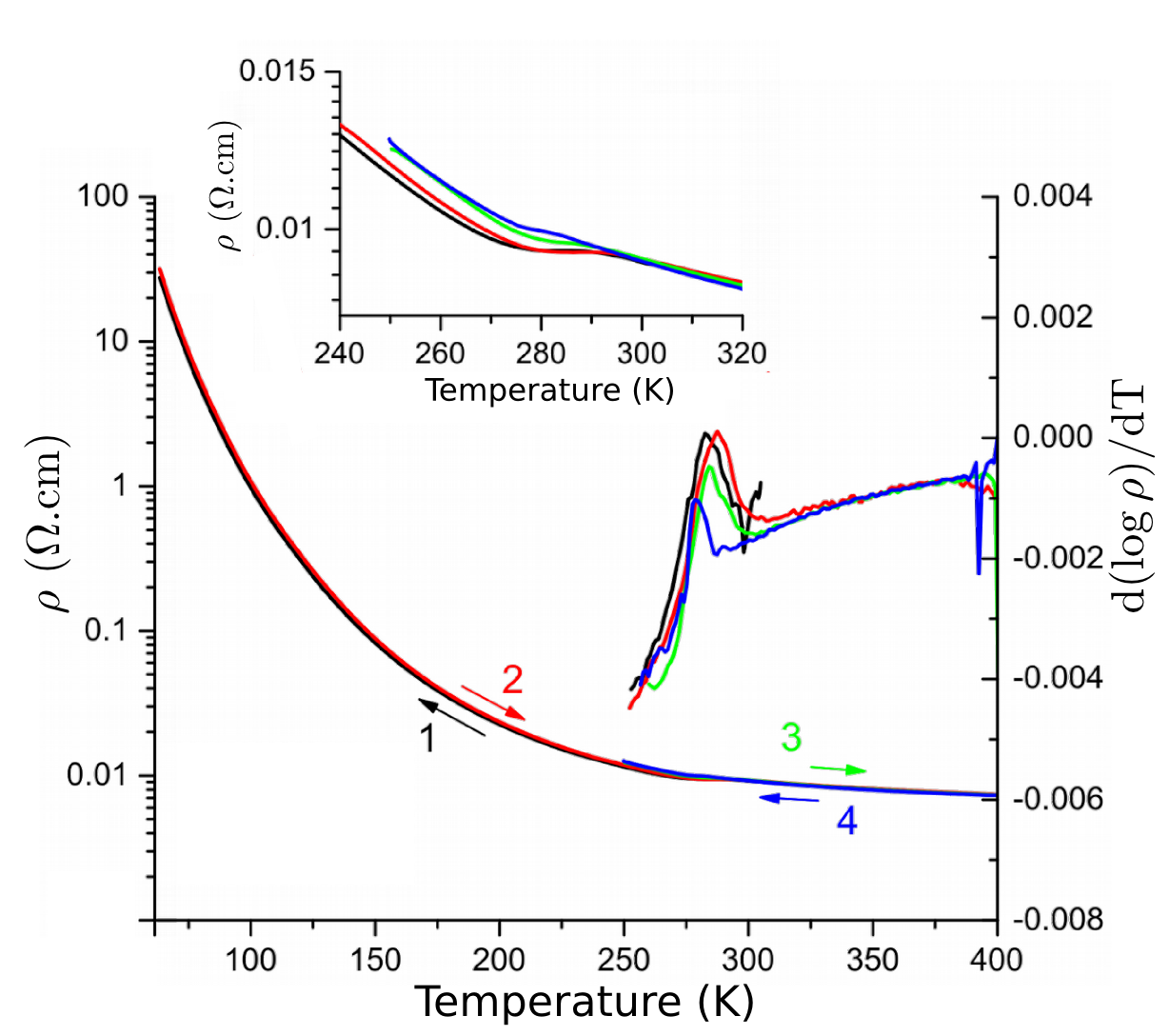}
  \caption{\textbf{Temperature dependence of the electrical resistivity} 
Electrical resistivity of Bi$_{0.35}$Sr$_{0.18}$Ca$_{0.47}$MnO$_{3}$ as a function of temperature and thermal cycling.
As shown in the inset, a resistivity anomaly associated with charge-ordering occurs just above room temperature ($\sim300K$) and has thermal cycling dependence (color and arrows indicate heating or cooling).
The charge-ordering critical temperature is seen more clearly in the $\frac{d(\log\rho)}{dt}$ plot.
Instead of a sharp phase transition, we observe a broad and gradual transition.
Note that resistivity measurements are performed on single crystals ($\sim$0.5 $\times$ 0.5 mm) with multiple crystalline domains (typical size $\sim$10-100$\mu$m).
}
  \label{F:Transport}
\end{figure}

\clearpage

\begin{figure}
  \includegraphics[width=6in]{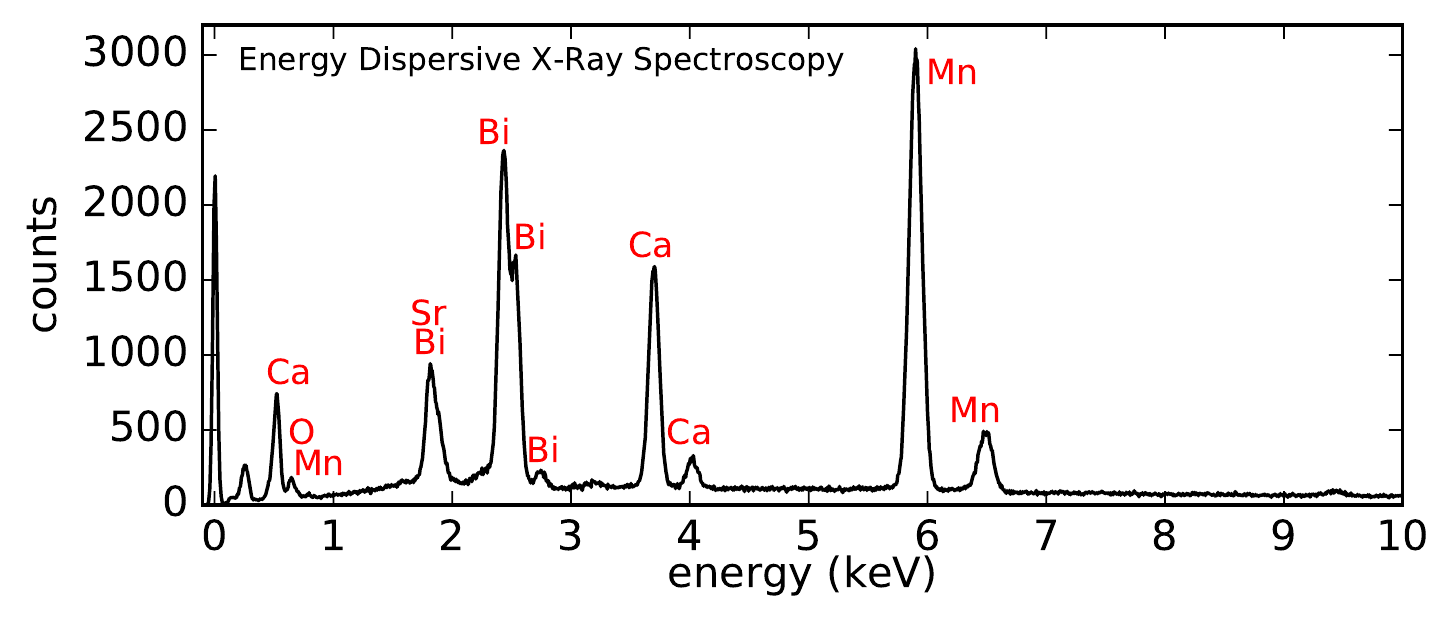}
  \caption{\textbf{Energy Dispersive X-ray Spectroscopy} 
The sample composition was determined using energy dispersive X-ray spectroscopy. We determined the Bi$_{1-x}$Sr$_{x-y}$Ca$_{y}$MnO$_{3}$ composition to be approximately $x=0.65$ and $y=0.47$.  We observed negligible variations in the composition across the sample.}
  \label{F:edx}
\end{figure}

\clearpage

\clearpage

\clearpage

\clearpage

\clearpage

\clearpage

\clearpage

\clearpage

\clearpage

\clearpage

\begin{figure}
  \includegraphics[width=5.5in]{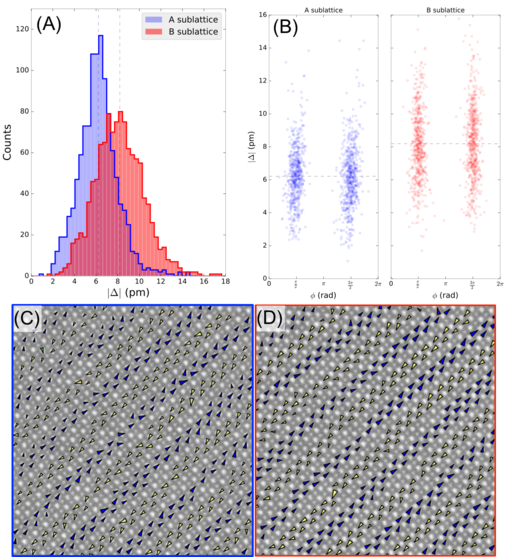}
  \caption{\textbf{A and B sublattice amplitudes} 
(\textbf{A}) Histograms of the displacement magnitudes at the PLD maxima in well-ordered regions on the A-sublattice (\textit{blue}) and B-sublattice (\textit{red}).  
The mean magnitudes at these sites is 6.2 pm / 8.2 pm on the A- / B-sublattices, respectively (\textit{dashed lines}).
(\textbf{B}) Scatterplots of the PLD displacement magnitude versus polarization angle $\phi$ for the A- and B-sublattices.  
The dashed lines again show the mean displacements at the sites analyzed.
(\textbf{C,D}) Sample sites used in analyzing the displacement magnitudes on the A- and B-sublattice are shown in C and D, respectively. 
To extract meaningful values we examined only sites corresponding to local PLD maxima.}
  \label{F:sublattice_amplitudes}
\end{figure}

\clearpage

\clearpage

\begin{figure}
  \includegraphics[width=6in]{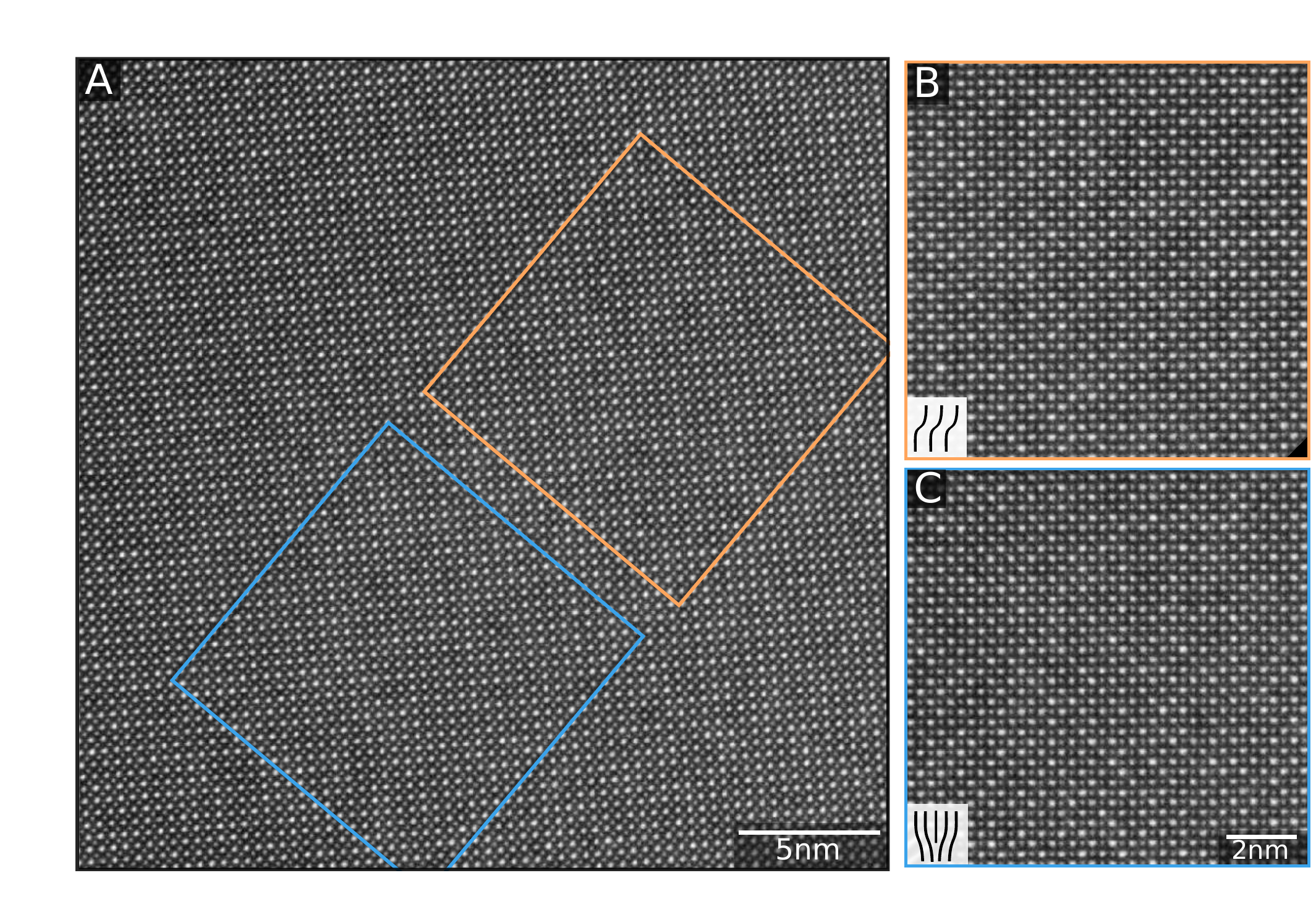}
  \caption{\textbf{HAADF lattice image of BSCMO} 
(\textbf{A}) Original, unrotated HAADF image corresponding to results in main text.
(\textbf{B}) Raw data corresponding to the region containing the shear deformation shown in Figs.~4A-C in the main text.
(\textbf{C}) Raw data corresponding to the region containing the dislocation shown in Figs.~4D-F in the main text.
No dislocations are observed in the underlying lattice, supporting that observed PLD defects in main text are intrinsic to the modulations.
HAADF data are unprocessed except for registration and alignment of image series (see \textit{Materials and Methods}).
The A-sites exhibit varying intensities indicating quenched impurity disorder due to Bi/Sr/Ca doping. 
}
\label{F:UnprocData}
\end{figure}

\clearpage

%%%% Bibliography %%%%